\documentclass[aip,prb,amsmath,amssymb,reprint]{revtex4}
\usepackage[dvips]{graphicx}
\usepackage{dcolumn}
\usepackage{color}
\usepackage{ulem}
\usepackage{mathptmx, helvet, courier}
\usepackage{bm,amsmath,amssymb}
\usepackage{threeparttable}
\usepackage{multirow}
\usepackage{array}
\usepackage{etoolbox}

\usepackage{titlesec}
\titleformat{\paragraph}
{\normalfont\normalsize\bfseries}{\theparagraph}{1em}{}
\titlespacing*{\paragraph}
{0pt}{3.25ex plus 1ex minus .2ex}{1.5ex plus .2ex}

\begin{document}

\title{Structural changes in barley protein LTP1 isoforms
at air-water interfaces}

\author{Yani Zhao}
\affiliation{Institute of Physics, Polish Academy of Sciences,
 Al. Lotnik{\'o}w 32/46, 02-668 Warsaw, \\ Poland}

\author{Marek Cieplak}
\email{mc@ifpan.edu.pl}
\affiliation{Institute of Physics, Polish Academy of Sciences,
 Al. Lotnik{\'o}w 32/46, 02-668 Warsaw, \\ Poland}

\date{\today}

\begin{abstract}
\noindent
We use a coarse-grained model to
study the conformational changes in two barley proteins, LTP1 and
its ligand adduct isoform LTP1b, that result from their adsorption to
the air-water interface. The model introduces the interface
through hydropathy indices. We justify the model by all-atom simulations.
 The choice of the proteins is motivated 
by making attempts to understand formation and stability of
foam in beer. We demonstrate that both proteins flatten out at the interface
and can make a continuous stabilizing and denser film.
We show that the degree of the flattening depends on the protein --
the layers of LTP1b should be denser than those of LTP1  --
and on the presence of glycation.
It also depends on the number ($\le 4$) of the disulfide bonds in the proteins.
The geometry of the proteins is sensitive to the specificity of 
the absent bonds. We provide estimates of the volume of cavities
of the proteins when away from the interface.
\end{abstract}

\maketitle

\section{Introduction}

Proteins are usually studied in the environment of bulk water but there are
many situations where they can be found at interfaces of various kinds, such
as between water and solids (see, e.g. refs.\cite{bhakta2015,Nawrocki}), water
and organic microfibers (see, e.g. ref.\cite{Nawrocki1}), water and 
oil \cite{Sengupta}, or between water and air \cite{Leiske}.
An interface between air and water may trap proteins because of 
their heterogeneous sequential composition: the hydrophilic amino acid
residues tend to point towards water whereas the hydrophobic ones prefer to
avoid it. In particular, a system of proteins may form layers at the interface.
These layers have been demonstrated to exhibit intriguing viscoelastic 
and glassy properties \cite{Graham,Leheny,Murray,Leheny2}
that are of interest in physiology and food science.
For instance, the layers of lung surfactant proteins 
at the surface of the pulmonary fluid generate defence mechanisms
against inhaled pathogens \cite{Head} and provide stabilization
of alveoli against collapse \cite{lungs}.
Protein films in saliva increase its retention and facilitate
its functioning on surfaces of oral mucosa \cite{saliva}.
Adsorption at liquid interfaces has been demonstrated to lead to
conformational transformations in amyloid fibers \cite{Mezzenga}.\\

Here, we consider one interesting example of an air-water interface: foam
in beer, the character and abundance of which is considered to be
a sign of the quality of the beer itself. The foam forms by the rising
bubbles of CO$_2$ that form on openning the container.  
In our analysis,
the chemical differences between CO$_2$ and air are not relevant,  
because the interfacial behavior of proteins is introduced through a
simple consideration of hydropathy. However, in reality, the solubility 
of CO$_2$ in water is distinct from that of air \cite{solubility}  and
the pH factors of water depend on the gas dissolved.\\


Various proteins derived from malted barley, such as LTP1 (lipid transfer
protein 1) and protein Z, have been found to play a role in the formation
and stability of foam in beer \cite{beer,blasco2011}. In addition,
these proteins are quite special as they survive various stages of the
brewing process which involve heating and proteolysis. The purification
of LTP1 from beer through cation exchange chromatography 
has been found not to be well
separated from protein Z \cite{jegou2000}, suggesting existence
of some interactions between LTP1 and Z. 
Unlike LTP1, protein Z has been found to be resistant to the
digestion by protein A and to proteolysis during malting and
brewing \cite{leisegang2005}.\\

It is thus interesting to understand the properties of these proteins
and to explain their role in the foam formation by investigating
what happens to the beer proteins in a foam. We employ a coarse-grained
structure-based model of a protein \cite{Hoang2,JPCM,plos,Current} in combination with a
phenomenologically added force \cite{airwater} that couples to the hydropathy
index of a residue, in a way that depends on the distance to the center 
of the interface. The need to use such a simplified model, also involving
an implicit solvent, stems from the fact that atomic-level modeling
of an interface requires considering a huge system of molecules just to
maintain the necessary density gradient in the fluid. 
Placing proteins in it adds still another level of complexity related
to the long lasting processes of large conformational changes taking place
in the proteins. Nevertheless, we propose an all-atom model
to justify the coarse-grained approach qualitatively. In this model, the
interface is maintained by introducing a solid hydrophilic wall. The water
molecules stay near the wall leaving few molecules far away from it.
This setup generates an effective air-water interface which on an average,
is parallel to the solid wall.\\

\begin{figure}[h]
\centering
\includegraphics[width=0.15\textwidth]{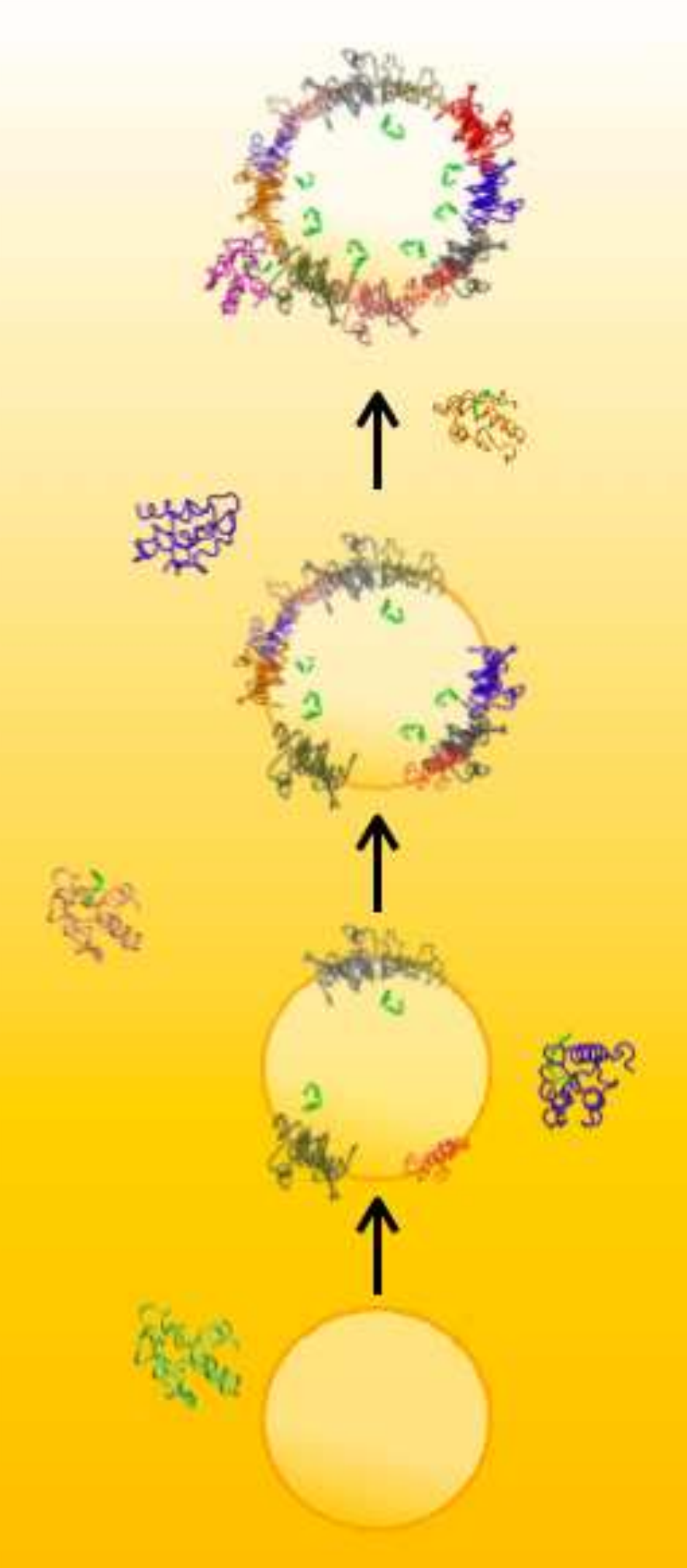}
\caption{Foam forms in beer by the rising CO$_2$ bubbles that occurs as a result of the reduction in pressure on opening the container. Those rising bubbles collect surface-active materials, such as LTP1 and LTP1b, which form an elastic skin around the bubble to stabilize it. This figure is the schematic representation of the adsorption of LTP1 and LTP1b to the surface of beer foams. The ligand of LTP1b is colored green. 
} \label{foamsche}
\end{figure}

We focus on protein LTP1 (PDB:1LIP) and on LTP1b (PDB:3GSH) -- its 
post-translationally modified isoform with a fatty ligand adduct.
We do not consider protein Z because its native structure is not known.
Proteins LTP1 and LTP1b are identical sequentially and their structures
differ by 1.95 {\AA} in RMSD (root-mean-square deviation) \cite{bakan2009}, 
 which is a measure of the average distance between atoms in
two superimposed conformations. As the reference conformation we take the 
crystal structure of the protein.
We study what happens to the conformations of these proteins
as they arrive at the air-water interface.
 The schematic representation of the adsorption of LTP1 and 
LTP1b to the surface of beer foams is shown in Fig.~\ref{foamsche}. 
These proteins form something like an elastic skin around a bubble and stabilize it. 
The ligand bound to LTP1b is ASY, which stands for
$\alpha$-ketol, 9-hydroxy-10-oxo-12($Z$)-octadecenoic acid \cite{bakan2009}.
The adduction process occurs during seed germination.
The ligand is formed from linoleic acid by the concerted action of 
9-lipoxygenase and allene oxide synthase \cite{bakan2009}.
It is known that 1 kg of the barley seeds produces 103.3 and 82.7 mg
of purified LTP1 and LTP1b proteins respectively \cite{jegou2000}.\\

Both isoforms contain four disulfide bonds when in barley 
seeds and in  malt. These  bonds form a cage delimiting 
the central hydrophobic cavity. In LTP1, the cavity 
is small but is capable of
capturing  different types of free lipids \cite{bakan2009}.
On the other hand, in LTP1b, the cavity much bigger but is filled
by about a half of the ASY ligand -- the other half is
outside of the protein.
With the use of the spaceball server \cite{spaceball}, we estimate that
the volume of the cavity, $V_c$, is 69.192 and 666.488 {\AA}$^3$ for LTP1 and 
LTP1b (on removing the ligand) respectively.
The sizes were determined by using a spherical probing
particle of radius 1.42 {\AA}, corresponding in size to the
molecule of water. The cavity in LTP1b actually consists of
three disjoint subcavities of comparable volumes.
It appears that the process of adduction makes the protein
looser and less rigid \cite{bakan2009,bettonigao}.
The near terminal regions move away from the center of the protein.
It should be noted, however, that the structure of LTP1
has been determined through the NMR method and that of LTP1b
by the X-ray crystallography.\\


During mashing, when hot water is added to malt to form wort
(obtained after the removal of insoluble fractions during lautering),
in which starches in barley are converted to sugars.
At this stage, also the LTP1 and LTP1b proteins get
glycated by forming covalent bonds with hexoses 
through the Maillard reaction \cite{jegou2001}.
Glycation increases hydrophilicity 
and solubility of LTP1 isoforms and
thus the foaming propensity \cite{jegou2000} without affecting
the structural stability of the protein \cite{reppocheau}.\\

In the next stage of beer production, wart boiling in the presence of a
reducing agent such as sodium sulphite, the disulfide bonds get
cleaved to varying degrees. The bigger the cleavage, the larger the 
probability for LTP1 and LTP1b to unfold irreversibly 
\cite{jegou2000,jegou2001} and thus to become better foam makers
because  the proteins flatten out wider at the interface and
thus generate a more continuous protein layer.
The native, compact isoforms display poor foaming properties \cite{bech1993}.
However, this simple picture gets complicated in the presence of
free lipids \cite{vannierop}. These lipids destabilise beer foam by 
disrupting the adsorbed protein layers at the interface 
into foam lamellae \cite{clark1994}. This destabilisation is significantly 
reduced by the presence of LTP1s, as these proteins capture the lipids 
into their cavities and thus reduce the free lipid concentration.
Thus, there should be an optimal degree of denaturation, at which LTP1s
are unfolded at the interface sufficiently well to generate an increased 
surface coverage and yet are still able to bind free lipids.
Interestingly, under the reducing conditions and in the presence of the 
lipids, LTP1b has been found to have a higher thermal stability than
LTP1 by 15$^\circ$C, as evidenced by the circular dichroism 
spectroscopy\cite{reppocheau}.\\

In our theoretical model, we include the ASY, sugar ligands, and
consider the proteins with various numbers, $n_{SS}$, of the disulfide bonds.
The disulfide bonds may get reduced during malting and brewing.
The reducing conditions are generated by malt
extracts and also by yeast\cite{reppocheau}.
We study what happens to the ligands and the geometry of the proteins
when they come to the interface. We show that the smaller the $n_{SS}$, 
the bigger the spreading of the proteins along the interface
and thus  larger stabilization of the foam. 
On the other hand, the thermal stability of the protein is expected 
to get reduced on lowering $n_{SS}$.
Our discussion of the geometry
also involves determination of the volume of the cavities
that LTP1 and LTP1b turn out to be endowed with (we study the case
of $n_{SS}$=4) and its dependence on the temperature, $T$. \\

It should be noted that the understanding of the properties of LTP1 is
interesting beyond just beer making. This protein
has been originally identified as promoting the transfer of lipids between 
donor and acceptor membranes in living plant cells \cite{carvalho}.
Its other physiological roles are not clear \cite{carvalho}.
However, LTP1 has been suggested to be important 
in the context of the response to changes in $T$ \cite{torres}, 
drought \cite{connell}, and bacterial and fungal pathogens \cite{michond}. 
LTP1 is known to act as an allergen in plant food, such as fruits, vegetables, 
nuts and cereals, latex and pollens of parietaria, ragweed, olive, and mugwort 
\cite{salcedo2007}.  \\      


\section{Methods}

In our coarse-grained model, we represent the LTP1 and LTP1b
proteins by 91 effective atoms placed
at the $\alpha$-C atoms. 
The interactions between the residues are described by the
Lennard-Jones potential of depth $\varepsilon$, approximately equal to
110 pN$\;${\AA} or 1.6 kcal/mol. The value of $\varepsilon$ is 
estimated by benchmarking simulations to the experimental results on the 
characteristic unraveling force for 38 proteins in bulk water \cite{plos}. 
 This value is consistent with what was derived through all-atom
simulations \cite{Poma}.
The length parameter, $\sigma$, is determined
from the native distance between the residues. These interactions
are assigned to native contacts, as determined through the overlaps
(the OV contact map \cite{Wolek}) between atoms belonging to the residues.  The cutoff of the
Lennard-Jones potential is 20 {\AA}. 
A contact is considered ruptured if the distance between the $\alpha$-Cs
exceeds 1.5 $\sigma$.
Pairs of residues that do not form a native contact interact through
steric avoidance. \\

\begin{figure}[h]
\centering
\includegraphics[width=0.48\textwidth]{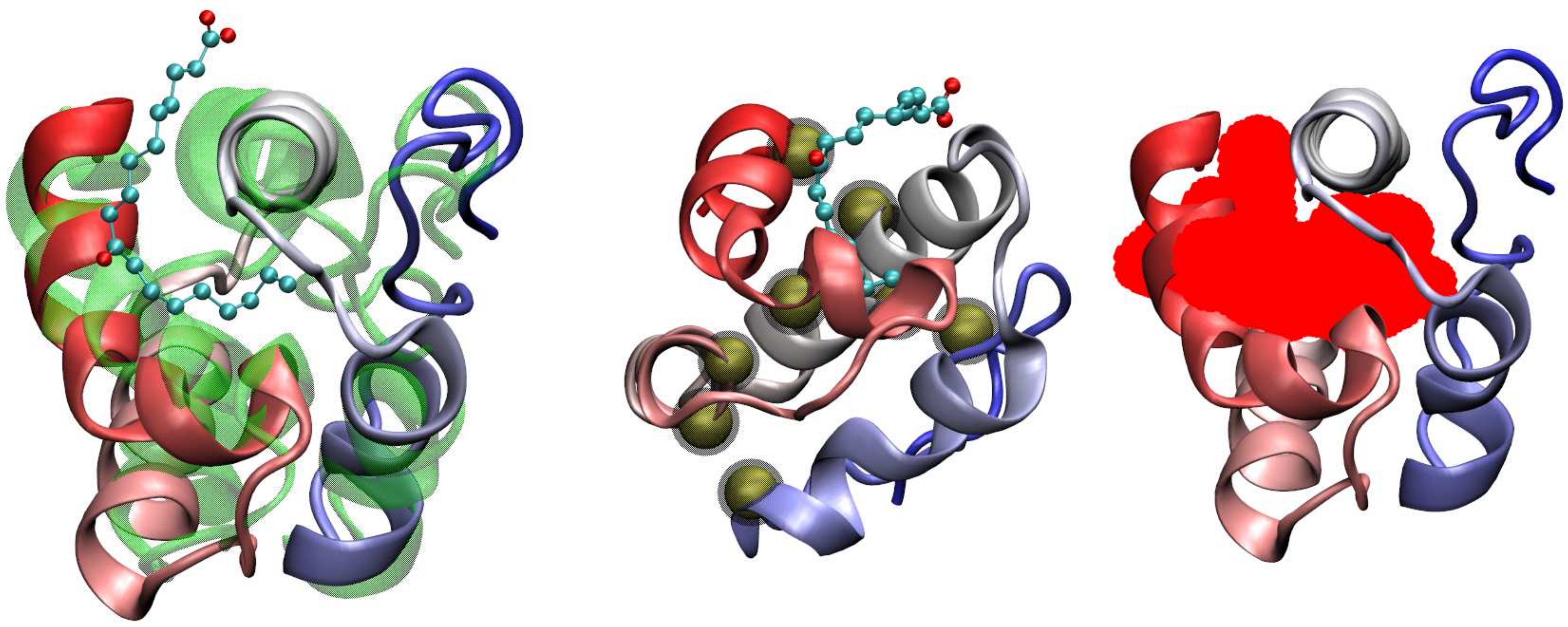}
\caption{Left: Superimposition of the crystal structures of LTP1 (green) and LTP1b 
(the protein is color-ramped from red to blue, from the N- to the C-terminus). 
The carbons and oxygens of the ligand of LTP1b are shown in cyan and red beads. 
Middle: The placement of the four disulfide bonds in LTP1b. The  eight cysteine 
residues involved in disulfide bonds are shown as black-yellow beads. The disulfide bonds in LTP1 connect the same sites as in LTP1b.
Right: The three cavities in LTP1b (on removing the ligand) are indicated in red.
} \label{cystalltp1}
\end{figure}

The four disulfide bonds connect cysteines at sites
\{3,50\}, \{13,27\}, \{28,73\} and \{48,87\}, as illustrated in
Fig.~\ref{cystalltp1}. They are described by the harmonic terms,
similar to the tethering interactions in the backbone.
Under the reducing conditions, some number of the disulfide bonds 
get cleaved and the properties of the resulting systems depend on the
identity of the bonds that stay. For instance, there are 6 ways of
removing two disulfide bonds. All possibilities are listed in
Table \ref{permuta}, together with the notation used.\\

The backbone stiffness is described by a chirality potential \cite{Wolek1}.
The solvent is implicit and is represented by the overdamped Langevin
thermostat. Most of the
molecular dynamics simulations are done at $T=0.3\; \varepsilon/k_B$
for which folding is optimal ($k_B$ is the Boltzmann constant); 
effectively, this corresponds to the room temperature. 
$T$ is controlled by introducing the implicit solvent as represented by
the Langevin noise and damping terms in the equations of motion
\begin{equation}
m\ddot{\textbf{r}} = -\gamma \dot{\textbf{r}}+F_c+\Gamma \;\;.
\end{equation}
Here, $F_c$ is the force due to all of the potentials that describe the
protein and $m$ is the mass of the residue.
We take the damping coefficient, $\gamma=\upsilon m/\tau$, where $\tau$ is a
characteristic time scale. $\tau$ is of order 1 ns which reflects the 
diffusional instead of ballistic nature of the motion in the implicit
solvent. The ballistic motion would correspond to the all-atom timescale
of order ps. The factor of $\upsilon$ in the expression for $\gamma$ controlls
the strength of damping. We have determined \cite{Hoang2} that $\upsilon \ge 2$
corresponds to overdamping when the inertial effects are minor. We take $\upsilon=2$
to have fast  overdamped dynamics.
$\Gamma$ is the random Gaussian force with dispersion $\sqrt{2\gamma k_BT}$
so that fluctuations are balanced by dissipation.
The Langevin equations of motion are integrated by using the fifth
order predictor-corrector scheme \cite{Tildesley}.\\ 

\begin{figure}[h]
\centering
\includegraphics[width=0.3\textwidth]{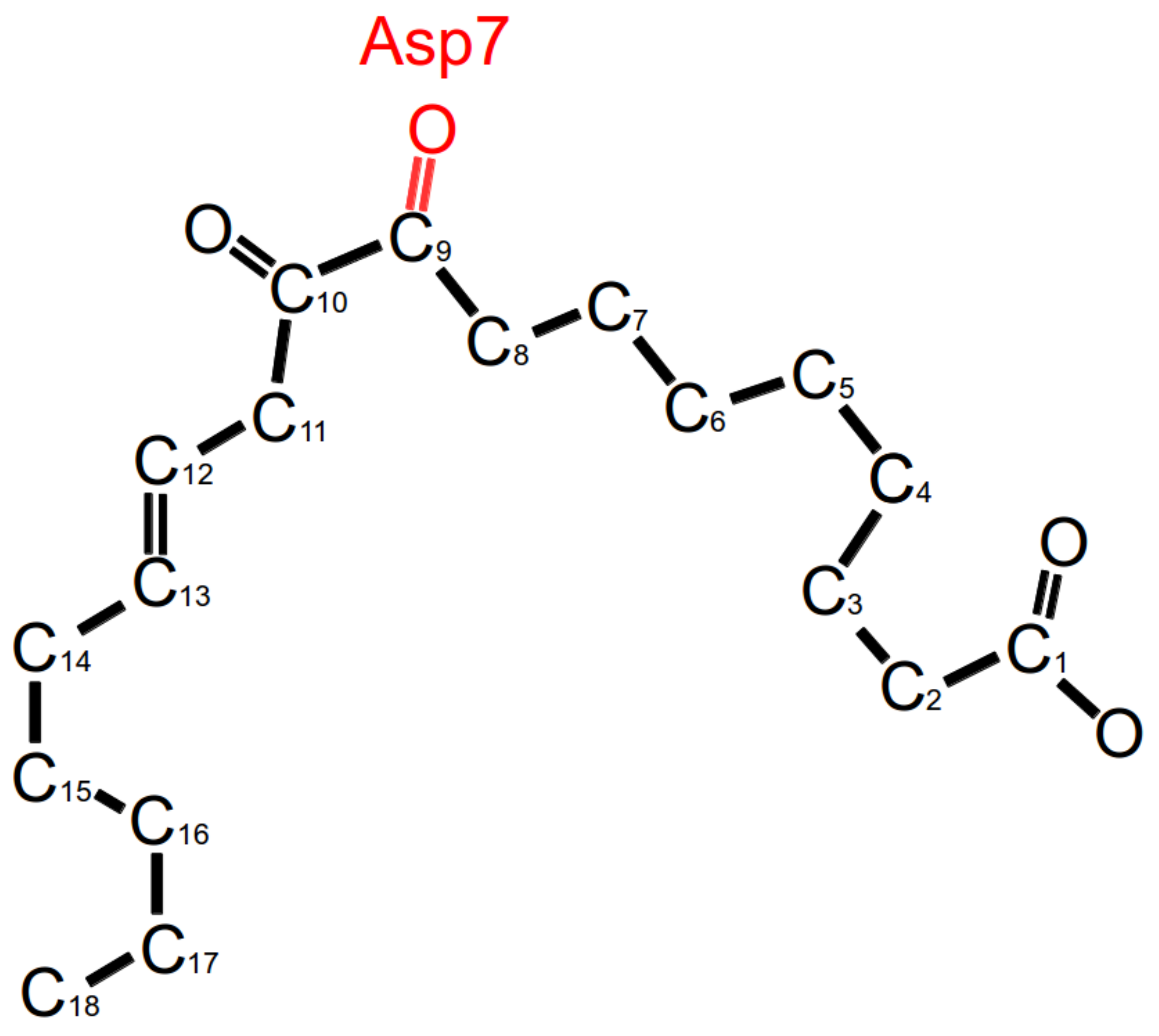}
\caption{Schematic representation of the atomic structure of ligand ASY.
The hydrogen atoms are not shown.
} \label{figlig}
\end{figure}

ASY consists of 18 carbon, 3 oxygen, and 32 hydrogen atoms. 
The 9th carbon C9 is covalently bound to the O2 atom of Asp 7 in protein LTP1b.
The bonding site splits the ligand into two branches, see  Fig.~\ref{figlig}. 
In the native state, the branch extending from C1 to C8 lies on the hydrophobic
surface of the protein, while the other branch is burried
within the cavity as shown in Fig.~\ref{cystalltp1} and \ref{figlig}. 
In the coarse-grained model, we represent ASY by 18 effective atoms
located at the carbon atoms. The native contacts between the residues
of LTP1b and ASY are determined by using the LPC/CSU server \cite{csu}.
There are 6 contacts that ASY makes with the outside part of the protein.
These are: C1--Gly53, C2--Gly53, C4--Ile54, C5--Gly57, C6--Gly57 and C8--Ile54.
In addition, there are 26 contacts within the cavity:
C10--Asp7, C10--Lys11, C11--Asp7, C11--Lys11, C11--Ile54, C12--Lys11, C12--Leu14, 
C12--Ile58, C13--Met10, C13--Leu14, C13--Ile58, C14--Met10, C14--Ile54, C14--Ile58,
 C15--Met10, C15--Val17, C15--Ile58, C16--Met10, C16--Leu51, C17--Ile54, C17--Ala55, 
C17--Ile81, C18--Ala55, C18--Leu61, C18--Ala66 and C18--Ile81.\\

In our theoretical study, the presence of the air-water interface is simulated 
by an interface-related force that is coupled to the hydropathy index, $q_i$, 
of the $i$th  residue \cite{airwater,yani2016}.
It  is given by 
\begin{equation}
F_i^{wa}=q_i\;A \;\frac{\text{exp}(-z_i^2/2W^2)}{\sqrt{2\pi}W}
\label{hyforce}
\end{equation}
where $A$ is a measure of the strength of the force, $W$ is the width
of the interface, and $z$ is the Cartesian coordinate that measures
the distance away from the center of the interface. Generally, the negative
values of $z$ correspond to water and positive  to gas, but the 
transition between the two phases is gradual. 
The interface itself is in the $x-y$ plane. We use the values:
$A$=10$\; \varepsilon$ and $W=5\;${\AA}. They were selected so that
a protein arriving at the interface does not depart from it. 
The hydropathy indices are taken from ref. \cite{Kyte}. They range
from --4.5 for the polar arginine to 4.5 for the hydrophobic isoleucine.
This scale is close to that derived by Wolfenden {\it et al.} \cite{Wolfenden} as both
scales have been derived from the physicochemical properties
of amino acid side chains instead of from the probability of
finding a residue in the protein core, as done in refs. \cite{Janin,Rose}.
The force is acting on the hydrophilic residues points toward $z<0$
and on the hydrophobic ones toward $z>0$.
The overall hydrophobicity for a protein of $N$ residues is given by
$H\;=\;\frac{1}{N}\;\sum_{i=1}^N \; q_{i} $. For LTP1, $H$ is -0.38.\\

In the case of LTP1b, there is a need to define $q_i$ also for the
atoms of the ligand. For this purpose, we use the non-overlaping
molecular fragment approach \cite{clogp}, abbreviated as ClogP.
In this approach, one considers concentrations of a compound
that is present in two coexisting equilibrium phases of a system and
defines the partition coefficient as the ratio of these concentrations.
It is assumed that the coefficient for the compound can be estimated as
a sum of the coefficients of its non-overlapping molecular fragments.
The fragments consist of a group of atoms and the neighboring fragments
are assumed to be linked covalently. With the use of the
BioByte’s Bio-Loom program \cite{biobyte} we have determined that
the hydrophilic head of ASY, consisting of C1 and two oxygen
atoms (Fig.~\ref{figlig}), can be assigned the ClogP value of --0.5.
The tail C2--C18 is hydrophobic and each of the carbons in the tail 
has the ClogP value of 0.35. These ClogP values are taken the
estimates of $q_i$. For alanine, this approach yields 1.1 which
is close to the Kyte and Doolittle value of 1.8 \cite{Kyte}.
For LTP1b, we get $H=-0.27$.\\

\subsection{Justification of the phenomenological model of the interface}

 In order to provide a qualitative atomic-level justification of the
hydropathy-based model, we use the NAMD \cite{NAMD} all-atom molecular dynamics
package with the CHARMM22 force field \cite{charmm1,charmm2} and consider the following 
simulation set-up. The system is placed in a box which extends between --50 and +50 {\AA}
both in the $x$ and $y$ directions. In the $z$ direction, it extends between --90
and +90 {\AA}. We situate the center of mass of LTP1 at point (0,0,0) and
freeze the protein in its native state. We consider LTP1 with $n_{SS}$=4 and 0.
There is a multitude of possible orientations
that the protein can make. We select two which are
defined with respect to the direction of the hydropathy vector \cite{airwater}.
This vector is defined as $\vec{h} = \frac{1}{N} \sum_{i=1}^{N}q_i \vec{\delta_i}$, 
where $\vec{\delta_i}$ is the position vector of the $i$th residue with respect 
to the center of mass of the protein. Orientation I corresponds to $\vec{h}$
pointing towards the positive $z$ axis. One of the hydrophobic residues 
at the top is leucine-61 and one of the hydrophilic residues at the
bottom is glutamine-39. The distance between the $\alpha$-C atoms of these
residues will be denoted by $d_h$. Orientation II is when $\vec{h}$ points
in the opposite direction: leucine-61 is at the bottom and glutamine-39
at the top.\\

\begin{figure}[h]
\centering
\includegraphics[width=0.5\textwidth]{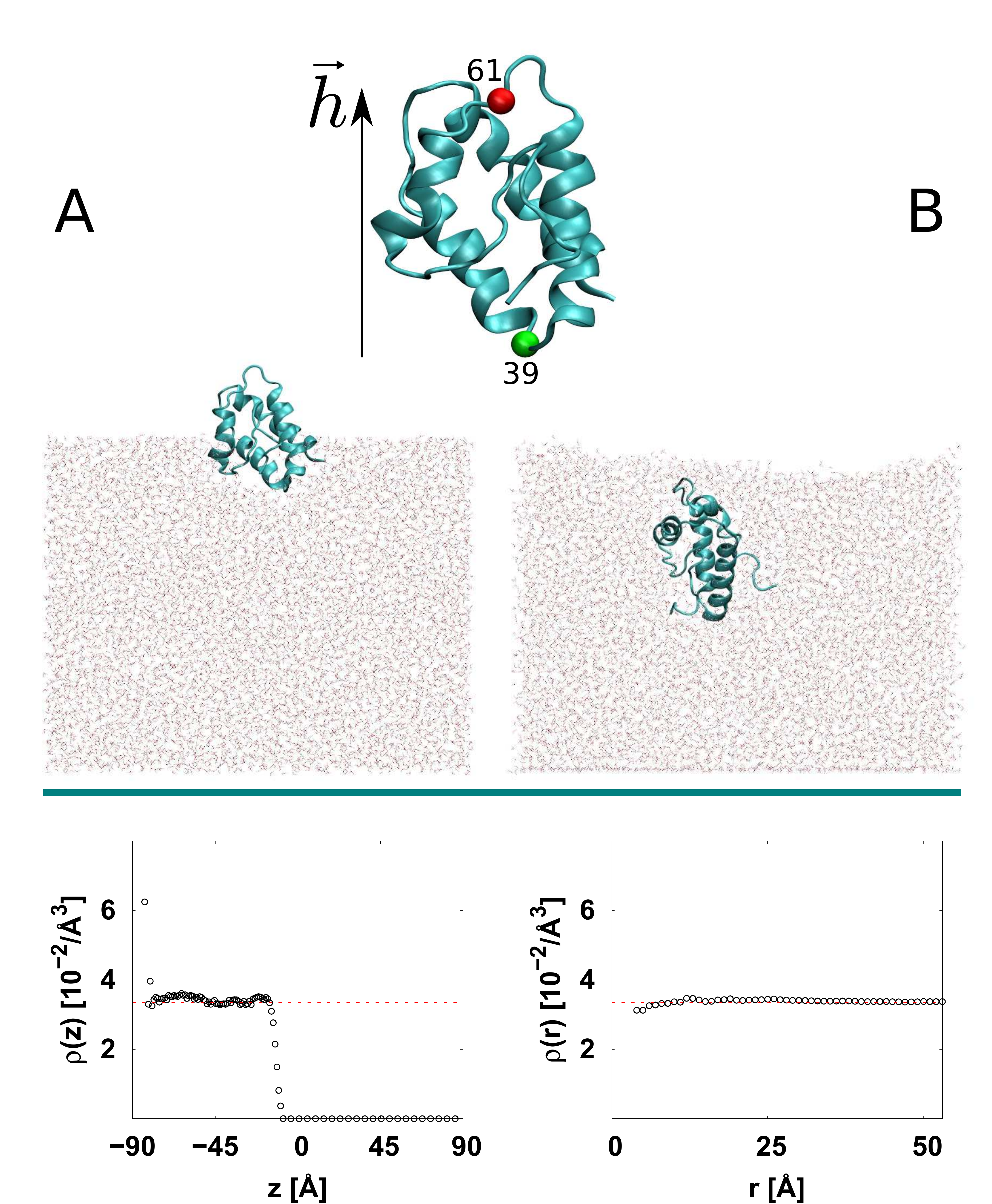}
\caption{Top: Protein LTP1 in orientation I. The red sphere corresponds
to the hydrophobic leucine-61 and the green sphere to the hydrophilic
glutamine-39.
Middle: The left panel shows the initial placement of the protein in water
in orientatin I. The right panel shows LTP1 with $n_{SS}=4$ 12 ns later.
A wall composed by asparagine residues is placed at the bottom of the system.
The water molecules displayed are from the slice between $y$=--13 and 20  {\AA}
(the $y$ direction is perpendicular to the plane of the figure).
Bottom: The number density profiles of the air-water system along $z$-axis and 
the radial direction in the $x-y$ plane. The red dashed line indicates
the level characterizing water under atmospheric pressure.
} \label{namdinitial}
\end{figure}

We then place 26 381 molecules of water, as described by the TIP3P model
\cite{water}, in the space corresponding to $z\le 0$ and outside of the
region occupied by the protein. Two Cl$^-$ ions are added to the solvent
to neutralize the charge of LTP1. We do not build a specific ionic strength
because it is not clear what it should be. In order for the water molecules
to prefer staying in the lower half of the simulation box,
we set a hydrophilic wall at $z$=--90 {\AA}. The wall is made of a
single layer of 6728 asparagines (see panel A of Fig.~\ref{namdinitial}). The
$\alpha$-C atoms of the asparagines ($q_i$ of -3.6)  are anchored to the sites of 
the [001] face of the fcc lattice with the lattice constant of 5 {\AA}. The side 
groups of the residues are directed towards water and they stay frozen. We use 
periodic boundary conditions and the Particle Mesh Ewald method \cite{Darden}.
\\

The system of the water molecules is then equilibrated at $T=$300 K for 2 ns. 
The number density profile of the water molecules along $z-$ and
the radial direction in the $x-y$ plane  is shown in the bottom 
panels of Fig.~\ref{namdinitial}. The results are time-averaged over
5 ns based on frames as obtained every 20 ps with the protein staying frozen.
The density profile $\rho(r)$ in the $x-y$ plane is averaged
over all values of $z$ except for the immediate vicinity of the bottom wall.
It is  calculated starting from $r=5$ {\AA} to avoid the excluding effects of 
the protein.
The attractive wall pulls water in
and sets the number density of water at $3.37 \pm 0.60 \times 10^{28}$ m$^{-3}$
which is consistent with $3.34 \times 10^{28}$ m$^{-3}$ for water under
normal conditions. We observe  that $\rho(z)$ goes down from the bulk value
to zero at $z=-5$ {\AA} and the width of the interface is about 8 {\AA}.
The radial distribution function, averaged over the
regions of bulk water, is nearly constant.\\

We now unfreeze the protein and equilibrate the whole system in two steps:
2 ns at $T$=150 K and 10 ns at 300 K. The protein is overall hydrophilic
so it gets drowned in water, as shown in panel B of Fig.~\ref{namdinitial} 
for the case of $n_{SS}=4$, but it stays at the interface for about 11 ns.
The water coverage
in the panel is shown in an exaggerated way because all molecules in the
system are projected into the $x-z$ plane.\\

\begin{figure}[h]
\centering
\includegraphics[width=0.5\textwidth]{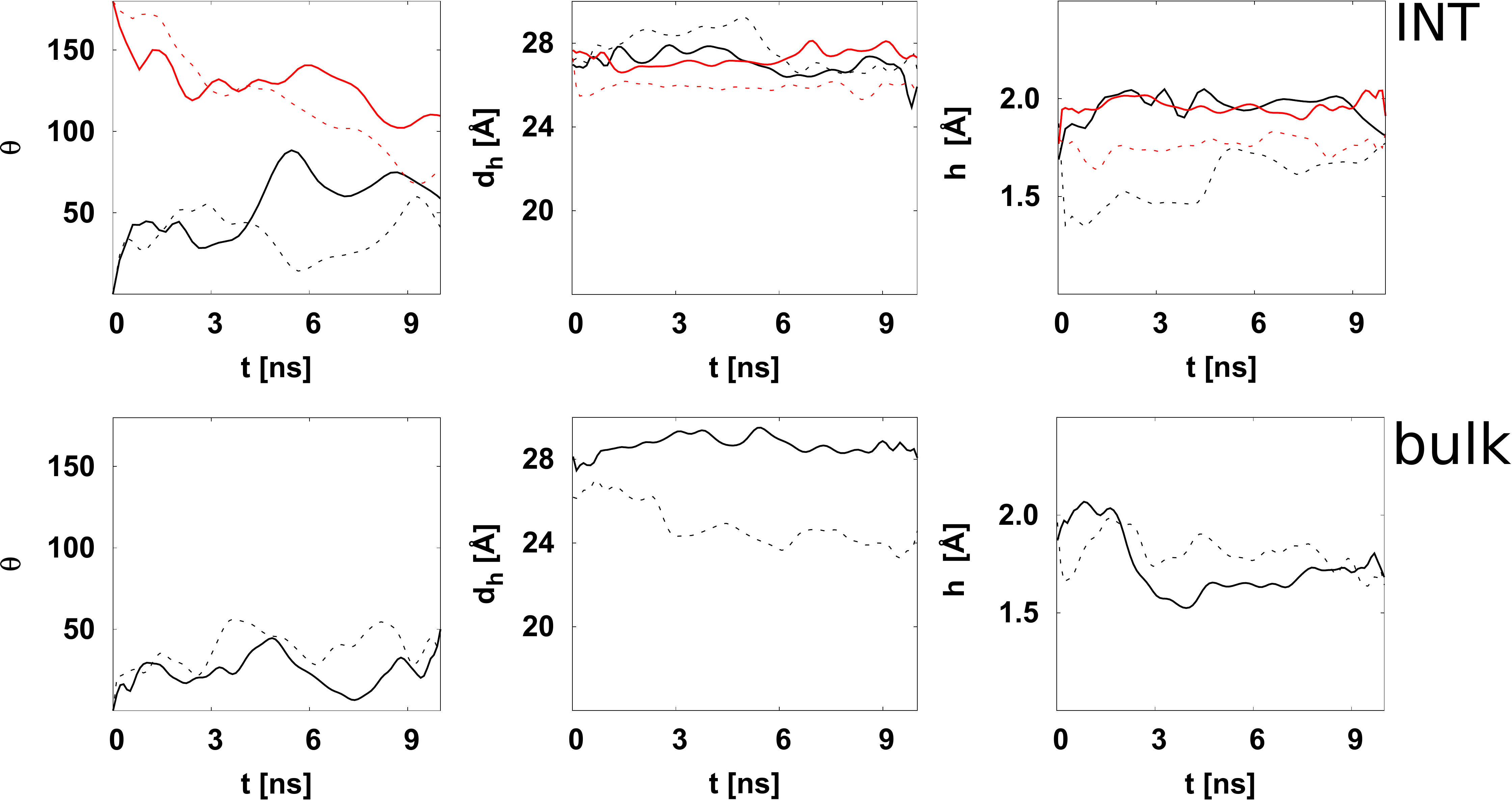}
\caption{Comparison of the behavior of the protein at the interface and 
in the bulk water as characterized by $\theta$, $d_h$, and $h$. 
The black lines are for orientation I: solid line for $n_{SS}$=4 and
the dashed line for $n_{SS}$=0. The red lines are for orientation II
with the similar convention regarding the value of $n_{SS}$.
In the bulk situation, the initial orientation is selected randomly.
} \label{namdrotate}
\end{figure}

We monitor the orientation and the change in shape of the protein in
the time interval in which it is pinned at the interface.
One parameter is $\theta$ - the angle that the vector $\vec{h}$ makes with the
$z$-axis. Initially it is 0 for orientation I and 180 for orientation II.
In bulk water, $\theta$ is measured with respect to the initial
random orientation. At the interface,
orientation I should favor not making any major change in $\theta$.
Instead, it evolves to about 70$^\circ$, as shown in the left panels 
of Fig.~\ref{namdrotate}. The reason is that the
force field we use is not fully compatible with the hydropathy indices
-- the indices have not been obtained through molecular dynamics calculations.
We observe that when one starts with the 70$^\circ$ orientation then the
protein just fluctuates around it for as long as it stays at the interface,
indicating a compatibility with the hydropathy related forces.
Thus molecular dynamics may provide a way to rederive hydropathy indices.
For orientation II, it is expected that $\theta$ would merge with
the range of values obtained for orientation I if it could stay at the
interface longer.\\

In order to monitor the changes in the shape, we consider $d_h$ and $h=|\vec{h}|$.
For $n_{SS}$=4 both parameters are close to that obtained in bulk water
(the middle and right panels in Fig.~\ref{namdrotate}).
However, for $n_{SS}$=0 both parameters indicate an expansion compared
to the bulk situation.
We conclude that the atomic-level considerations support the
orientational and conformational effects produced by the phenomenological 
model described by Eq. (\ref{hyforce}). Events of the interface
depinning (often followed by events of repinning)
can by captured by a reduction in the value of $A$.
However, our test runs do not indicate any depinning for several
other proteins. Other force-fields may extend the time at the interface
for LTP1. We work in the limit in which no depinning is expected.\\

\section{Results}

\subsection{Properties of single proteins away from the interface}

We characterize the equilibrium properties of the proteins by three quantities:
$P_0$, $Q$, and the root mean square fluctuation (RMSF), which is 
a measure of positional fluctuations of a residue with respect to its initial
location. These quantities are determined
based on 5 long runs 
(100 000 $\tau$ each) that start in the 
native state and correspond to a temperature $T$.
The first of these is the probability of all native contacts being present
simultaneusly.  The disulfide bonds do not count as contacts.
The  temperature, $T_0$, at which $P_0$ crosses through $\frac{1}{2}$
is a measure of the melting temperature. $Q$, on the other hand, is a 
fraction of the native contacts that are established (i.e. without the
condition of the simultaneous presence of all contacts). The temperature, $T_Q$, at which
$Q$ crosses through $\frac{1}{2}$ is close to a maximum in the specific
heat which signifies a transition between extended and globular conformations.
This point is discussed further in ref. \cite{Wolek1}. $T_Q$ is necessarily
much higher than $T_0$. $P_0$ and $Q$ provide global characterization whereas 
RMSF give local information about the magnitude of the positional
fluctuations of the $i$th residue.\\

\begin{figure}[h]
\centering
\includegraphics[width=0.45\textwidth]{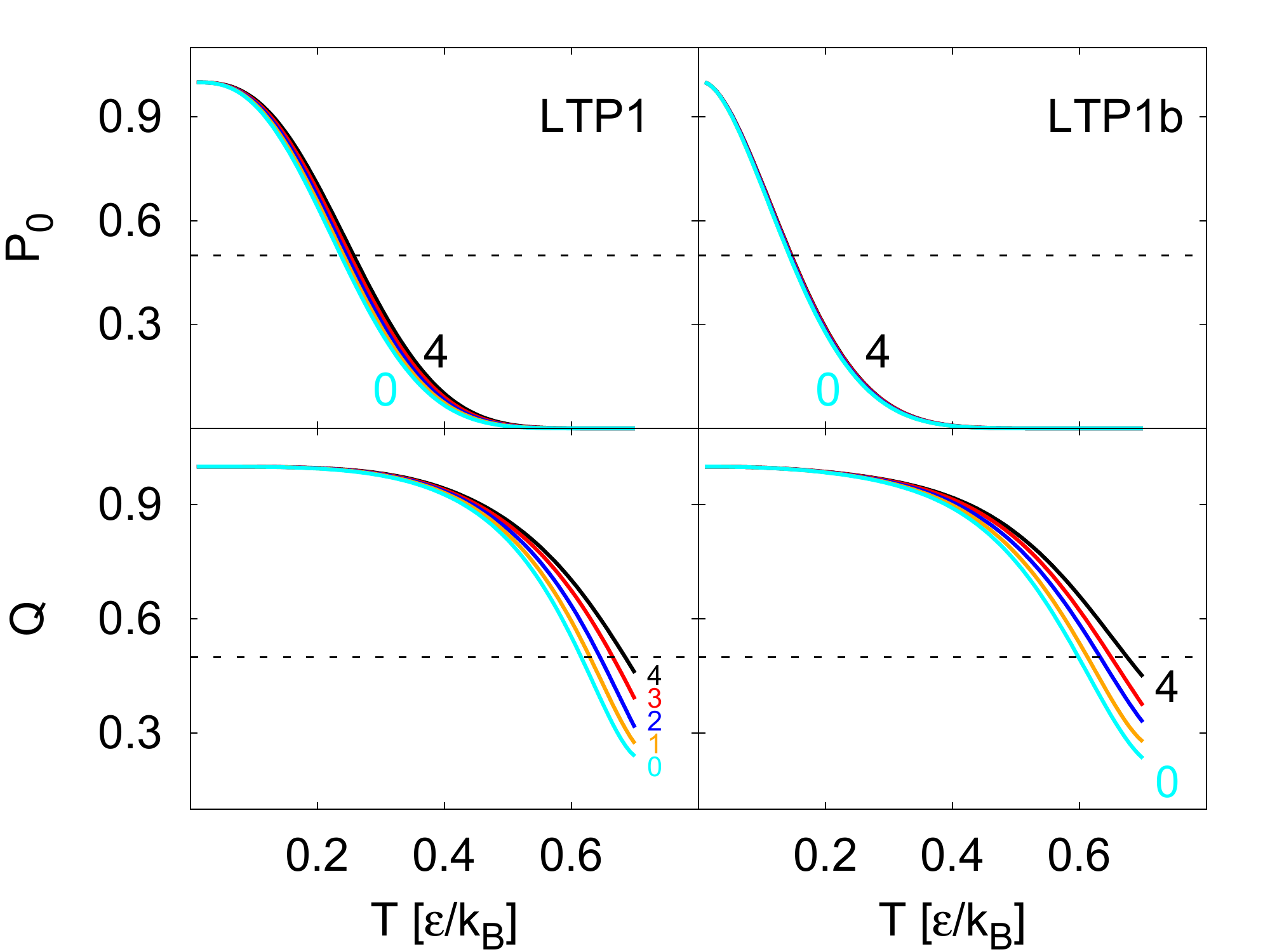}
\caption{The temperature dependence of $P_0$ and $Q$ for LTP1 (left) 
and LTP1b (right) in bulk water for the indicated values of $n_{SS}$. 
The data points are averaged over all possible permutations
of the disulfide-bond placement.
The dashed lines correspond to the level of $\frac{1}{2}$.
} \label{QP}
\end{figure}

Fig. \ref{QP} shows the $T$-dependence of $P_0$ and $Q$ for LTP1
for various values of $n_{SS}$. The data is averaged over the permutations. 
The values of $T_0$ do not vary much: it is 0.257 $\varepsilon/k_B$ for
$n_{SS}$=4 and 0.236 $\varepsilon/k_B$ for $n_{SS}$=0 -- less than 1$^\circ$C
difference.
At $T$=0.5$\;\varepsilon/k_B$, i.e. at about 100$^\circ$C, the differences in
$Q$ remain small, indicating a remarkable thermal stability.
Our observations agree with the results in ref.~\cite{larsen2001}
that there is no major structural change in LTP1 taking place 
between 20 to 90$^{\circ}$C. 
For LTP1b, the ligand related contacts are counted in the calculation of 
$P_0$ and $Q$. These contacts are easy to rupture on heating, which 
results in LTP1b being a less stable structure than LTP1.\\

\begin{figure}[h]
\centering
\includegraphics[width=0.5\textwidth]{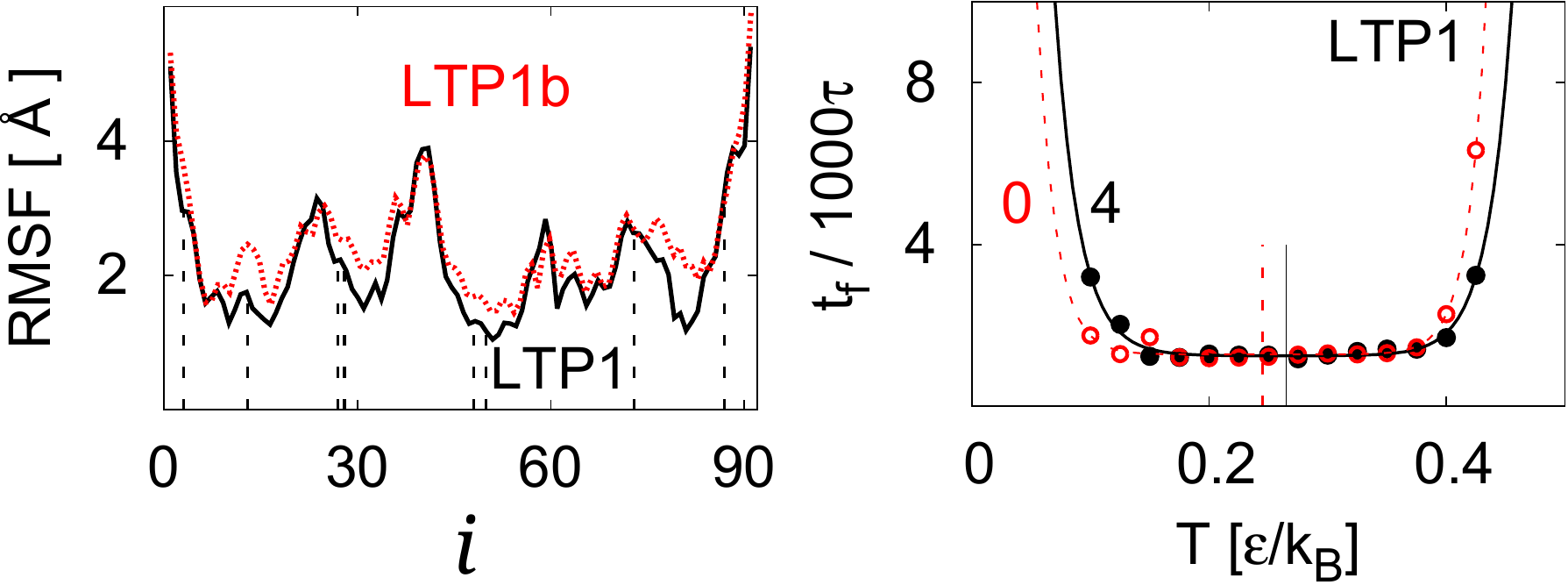}
\caption{
Left: RMSF of LTP1 (black lines) and LTP1b (dotted red lines) in bulk at 
$T=0.5 \varepsilon/k_B$ and for $n_{SS}=0$.
The RMSF of the eight cysteine residues involved in the formation of 
disulfide bonds are indicated by the dashed vertical lines.
Right: The temperature dependence of the median folding time, $t_f$,
for LTP1 with $n_{SS}$ equal to 4 (the black line) 
and 0 (the red line).
} \label{folding}
\end{figure}

Fig. \ref{folding} (the right panel) shows the $T$-dependence of the median
folding time, $t_f$, for LTP1 at $n_{SS}$=4 and 1.
$t_f$ is calculated by considering 100 trajectories which start
from a conformation without any contacts and by determining the median
time needed to establish all native contacts for the first time.
The dependence is
U-shaped and the center of the U defines the temperature of
optimal folding, $T_{min}$. We get $T_{min}$ of 0.26 and 0.24 $\varepsilon/k_B$ 
for $n_{SS}$ of 4 and 0 respectively, indicating an overall leftward
shift. The basins of good folding
are rather broad and $T_0$ is within the basins.\\

Fig. \ref{folding} (the left panel) compares the RMSF for LTP1 and LTP1b
at $n_{SS}$=0.
In order to enhance the difference between the patterns, the comparison
is done at $T$=0.5 $\varepsilon /k_B$.
The presence of the ligand is seen to increase the fluctuations at
almost all sites. The cysteine residues have varying levels of the RMSF.
We find that if a disulfide bond was replaced by a regular contact
then the most fragile of them is 3--50, and the most persistent is 48--87
(for LTP1, the probability of the contact being present is 56\% and 65\%
respectively; $T$=0.5 $\varepsilon/k_B$).
There are two reasons for the fragility of the 3--50 contact. 
First, residue 3 is close to the terminus. Second, the contact
has the largest contact order, as measured by the
sequential distance between the residues.
We conclude that, for both proteins, the 3--50 disulfide bond is the most
likely to be cleaved and 48-87 is the least likely.
Experimentally, one can study the reduction of disulfide bonds through
titration with DTNB (5,5'-Dithiobis-(2-nitrobenzoic acid) or Ellman's 
reagent), which reacts with sulfhydryl groups \cite{jegou2000}.
This method provides a reliable way to measure the concentration of
the reduced cysteines but it does not indicate the persistence levels
of specific bonds.\\

\subsection{Properties of single proteins at the interface}

\begin{figure}[h]
\centering
\includegraphics[width=0.5\textwidth]{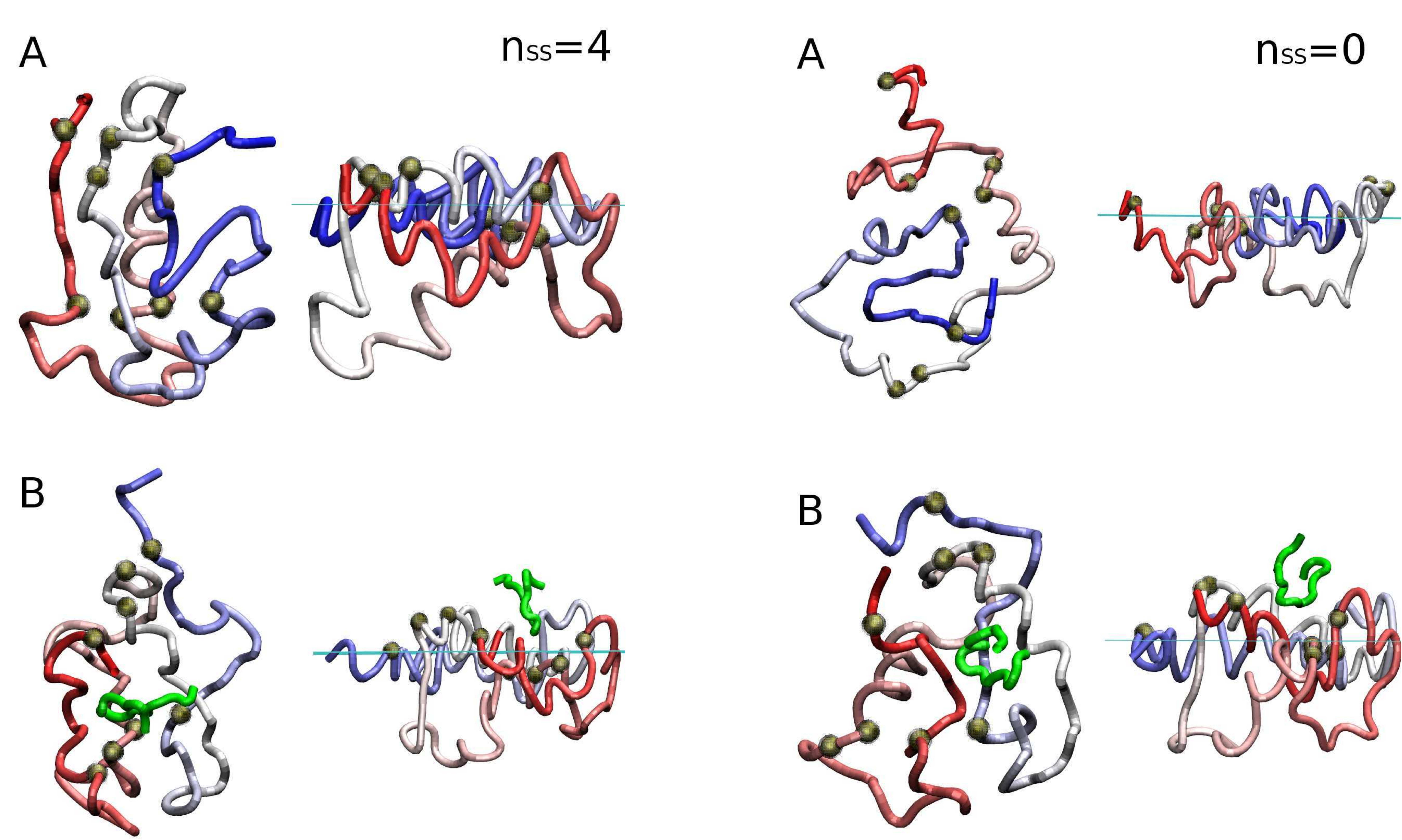}
\caption{Left: The top and side view of the most probable conformation of LTP1 (A) and LTP1b (B) at the air-water interface at $T=0.3 \varepsilon/k_B$ with $n_{SS}=4$. The 8 cysteines involved in disulfide bonds are shown in black-yellow beads and the ligand of LTP1b is shown in green. Right: The same as the left panel but for $n_{SS}=0$. The N-terminus of proteins is showed in red and the C-terminus is in blue.
} \label{peakx4}
\end{figure}

When studying the effects of the interface, we delimit the space by
repulsive walls at  $z$=-10 nm and $z$=10 nm and place the protein
close to the bottom wall, but still in "bulk water".
We then evolve the system for 10 000 $\tau$ to allow the protein to
come to the interface and to adjust to it.  The interface deforms
the protein, as illustrated in Fig. \ref{peakx4} for LTP1,
because of the hydropathy-related forces. In particular, we
observe the collapse of the cavity. In the case of LTP1b, the collapse
is concurrent with the expulsion of the ASY tails toward the gas phase.\\

In order to characterize the deformed geometry, we define three parameters:
$R_z$,  $d_z$, and $w$. The first of these is the radius of gyration
in the $x-y$ plane, 
$R_z=\{\frac1{2N^2}\sum_{i,j}\left[ (x_i-x_j)^2+(y_i-y_j)^2 \right]\}^{\frac{1}{2}}$,
where $x_i$, $x_j$, $y_i$ and $y_j$ are the $x$ and $y$ coordinates of 
the $i$ and $j$th residues. The second parameter is the vertical 
thickness of the protein, defined as the extension along the $z$-axis.
The third parameter describes the nature of the shape of
the protein. It is defined as 
$w=\Delta R/\bar{R}$ with $\Delta R=R_2-\bar{R}$ and $\bar{R}=1/2*(R_1+R_3)$.
$R_1$, $R_2$, and $R_3$ are the main radii of gyration, derived
from the moment of inertia, and ranked ordered from the smallest to
the largest. $w \approx 0$ corresponds to a globular shape,
$w<0$ to a flattened conformation, and $w>0$ to an elongated one. 
In the case of LTP1b, the ligand is taken into account in the calculation 
of the geometrical parameters.
When comparing a parameter $X$ (like $R_z$) between the two
proteins, we take LTP1 to be the reference system and define
the relative difference by
$r_X=(X_{\text{LTP1b}}-X_{\text{LTP1}})/|X_{\text{LTP1}}|$. \\

\begin{figure}[h]
\centering
\includegraphics[width=0.9\textwidth]{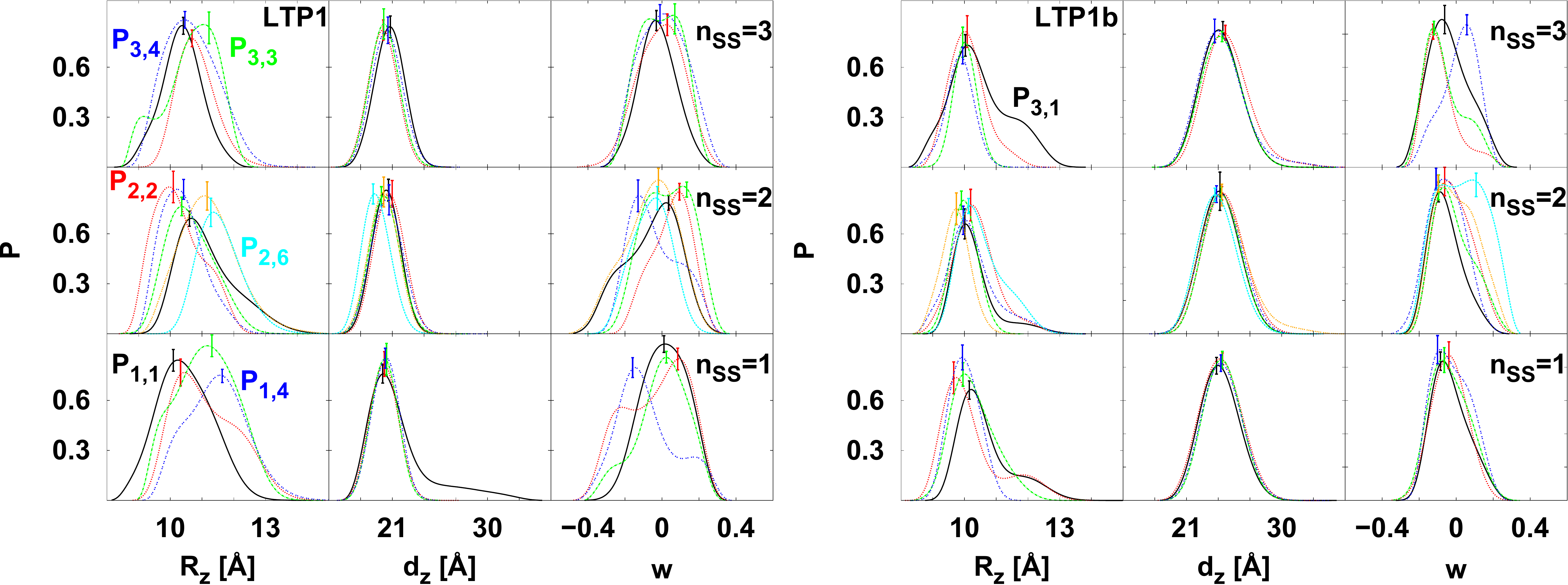}
\caption{The normalized histogram of $R_z$, $d_z$ and $w$ for LTP1 (left) 
and LTP1b (right) at the interface at  $T=0.3 \varepsilon/k_B$. 
The values of $n_{SS}$ range between 1 and 3.
In this plot, the permutation $P_{3,1}$, $P_{2,1}$ and $P_{1,1}$ are in black, 
$P_{3,2}$, $P_{2,2}$ and $P_{1,2}$ in red, $P_{3,3}$, $P_{2,3}$ and $P_{1,3}$
in green, $P_{3,4}$, $P_{2,4}$ and $P_{1,4}$ in blue, $P_{2,5}$ in orange,
and $P_{2,6}$ in cyan. 
} \label{seperS}
\end{figure}

\begin{figure}[h]
\centering
\includegraphics[width=0.4\textwidth]{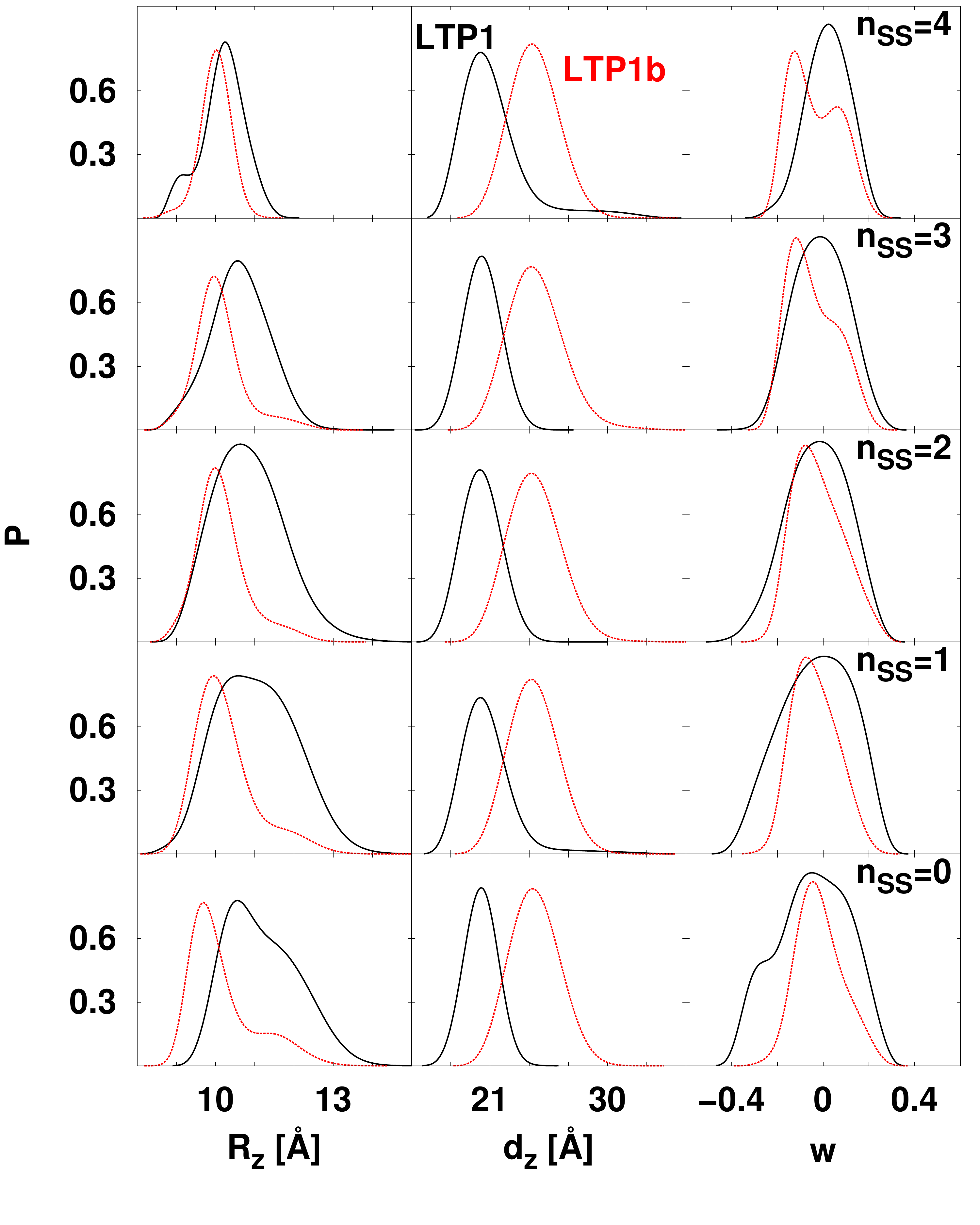}
\caption{The normalized histogram of $R_z$, $d_z$ and $w$ for LTP1 (black) 
and LTP1b (red) at the interface at $T=0.3 \varepsilon/k_B$.  
Tha data is for all possible values of $n_{SS}$ and for
all allowed permutations. 
} \label{hisrz}
\end{figure}

We generate 10 trajectories of coming to the interface analyze the conformations 
obtained at a permutation of $n_{SS}$.  Each trajectory lasts for 100 000 $\tau$
and we store the conformations obtained every 15 $\tau$ (1$\tau$ corresponds to
200 integration steps). In the analysis, we take into account only  those 
conformations in which the protein is at the interface.
Figs.~\ref{seperS} and \ref{hisrz} show the normalized histograms of 
$R_z$, $d_z$ and $w$ 
of partially reduced LTP1 and LTP1b 
at various stages of the disulfide-bond reduction.
Fig. \ref{seperS} distinguishes between the permutations
of the bond placement, if more than one is possible, whereas Fig. \ref{hisrz}
shows the distributions that are averaged over the permutations.
Table \ref{compinf} summarizes the results by averaging over the
distributions. Another summary is presented in Fig. \ref{shapex},
where the average values of $R_z$ and $d_z$, with the division
into the permutations and without, are plotted vs. $n_{SS}$. \\

Generally, the smaller the $n_{SS}$ the bigger the spread 
of the proteins in the $x-y$ plane.
One might expect that this effect should be coupled to the narrower
the vertical extension, but this is not necessarily so:
the dependence on the permutation dominates.
The average value of $w$ is close to zero but the spread in this
parameter is significant: between --0.4 and +0.4, indicating
a large variation in the shapes of the conformations. \\

We also observe a substantial sensitivity in the geometrical parameters
to the choice of the permutation.
For example, at $n_{SS}=2$, the most probable value (the highest peak in 
the histogram in Fig.~\ref{seperS}) of $R_z$ for LTP1 is at 10.04 {\AA} and it
is observed for permutation $P_{2,2}$. On the other hand, for permutation $P_{2,6}$
it is at 11.14 {\AA}. For $n_{SS}=1$, the difference between the most probable 
values of $R_z$s is about 1.6 {\AA} -- it is smaller for $P_{1,1}$ than for $P_{1,4}$.
The reason is that permutations $P_{2,2}$ and $P_{1,1}$ contain disulfide 
bond $\{3,50\}$ but exclude $\{48,87\}$, while $P_{2,6}$ and $P_{1,4}$ 
do the opposite. 
The disulfide bond $\{3,50\}$ involves sites that would fluctuate
more vigorously than $\{48,87\}$ if the bonds were broken. 
Thus permutations $P_{2,2}$ and $P_{1,1}$ limit 
the fluctuations in the protein maximally, which leads to  smaller 
values of $R_z$, than the other permutations.\\

In the case of LTP1b, the differences between various permutations
are smaller than for LTP1. The reason is that the dominant
shape-changing effect is due to the hydrophobic ligand which
induces stretching which is more vertical and comes with generally 
smaller values of $R_z$.\\
 
The distributions shown in Figs. \ref{seperS} and \ref{hisrz} are
mostly Gaussian but some display shoulders.
The shoulders reflect existence of various modes of the arrival
at the interface and, therefore, of different modes of action of the 
hydropathy-related forces. Examples of the conformations show in
Fig. \ref{peakx4} for $n_{SS}=4$ and 0 correspond to the dominant 
peaks in the distributions.\\

Fig. \ref{shapex} illustrates our observation that $<R_z>$ and, therefore,
the surface area of LTP1 and LTP1b increases as more disulfide bonds are 
cleaved. Moreover, the surface area of LTP1 is bigger than that of LTP1b
for  any value of $n_{SS}$ due to the vertical dragging action of the
ligand in LTP1b. Also, $<d_z>$ is by about 3 {\AA} larger for LTP1b than for LTP1.
We conclude that the reduced LTP1 protein is a better surfactant as
it spreads more and thus lowers the surface tension of foam.
However, LTP1b contributes to a better adsorption at the  interface,
packs better. Both kinds of proteins are present
in the foam and the two effects coexist.\\

\begin{figure}[h]
\centering
\includegraphics[width=0.6\textwidth]{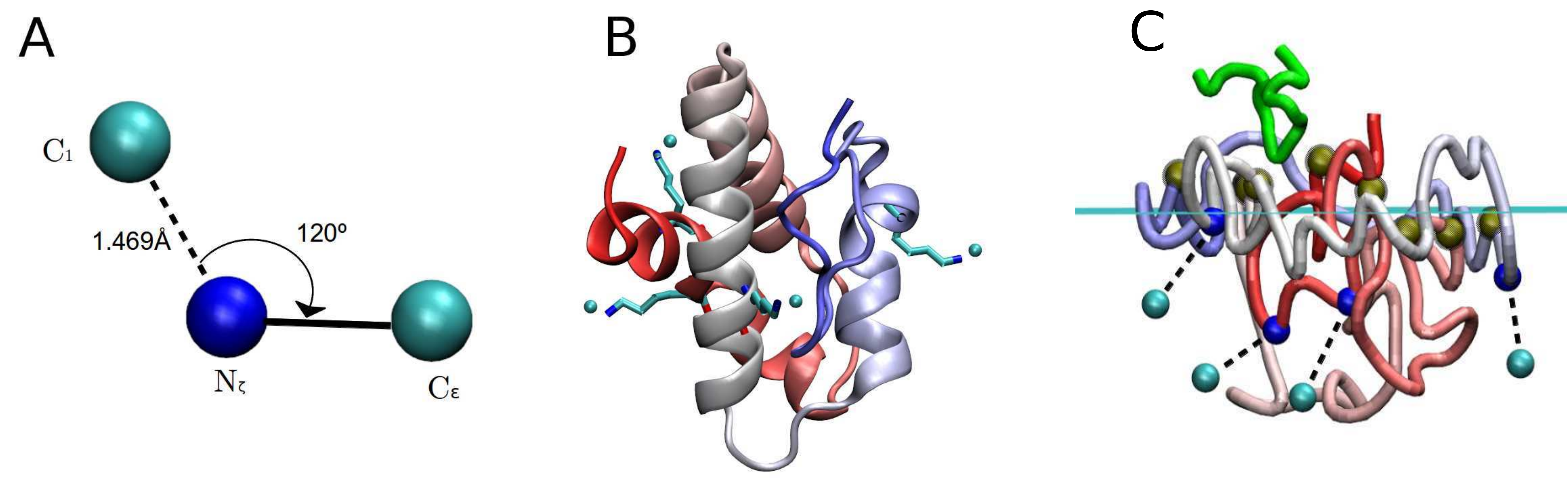}
\caption{A: Schematic representation of the C-N covalent bond (dotted line) 
formed between the C$_1$ atom of glucose and the nitrogen N$_\xi$ from the 
side chain of lysine. The bond length is 1.469 \AA, and the angle 
made by  C$_1$,  N$_\xi$ and the connected C$_\varepsilon$ is 120$^{\circ}$. 
B: Four glucoses (cyan) are covalently bound to four lysine residues of LTP1b
(the ASY ligand is not displayed). The N-terminal part of the protein is shown
in red and the C-terminal in blue.
C: The location of glucoses bound to  LTP1b at the interface. 
The glucoses and the lysine residues are shown as cyan and blue beads, 
respectively. The color coding of the terminal parts are as in panel B.
The 8 cysteines shown in dark-yellow beads and the ligand ASY is shown in green.
} \label{anglesuger}
\end{figure}

We now consider the role of glycation. It has been argued\cite{jegou2000}
that glycation involves the nitrogens either from the N-terminus
or from the nucleophilic amino
group on the side chain of lysine residues. It results in formation
of the C-N covalent linkage between the carbonyl group of sugars. 
Since the nitrogens on the side chains,  denoted by N$_\xi$, are more 
exposed in solution, we focus only on the glycation on lysins.
Concentration-wise, glucose and sucrose are the top two sugars
in barley after germination \cite{allosio2000}.
There are four lysins in LTP1 and LTP1b -- they can bind to four
sugar molecules each.
Here, we consider the case of glucose. The geometry of binding is 
illustrated in Fig.~\ref{anglesuger}.\\

In our coarse-grained model, each glycated glucose is represented as an
effective atom. The bond length of the C-N covalent bond is taken as 
1.469 {\AA}, \cite{bondl} and the bond angle as  120$^{\circ}$.
The rotation angle is taken randomly. The predicted ClogP value
for a glucose is -2.21 which signifies that it is hydrophilic.
As a result,  glucoses at the interface point toward bulk water, as illustrated
in Fig.~\ref{anglesuger}, panel C.\\

Fig. \ref{shapsuger} shows that glycation affects $<R_z>$ in a way that
depends on the level of reduction. For $n_{SS}$ equal to 0 or 1,
the surface area is enhanced  for LTP1, but is about the same for LTP1b.
For $n_{SS}$ of 2, it is enhanced for
LTP1 but decreased for LTP1b.
For $n_{SS}$ of
3 or 4, it is decreased for both proteins. 
The vertical spread $<d_z>$ is reduced on glycation
in all cases. We conclude that, at high levels of reduction, glycation
should enhance foam making by LTP1, but not by LTP1b.  
In beer, LTP1 and LTP1b coexist and the overall surface activity
of the system is thus enhanced, in agreement with ref. \cite{jegou2001}.\\

\subsection{Protein layers at the interface}

In order to study protein layers, we need to define the interactions
between individual proteins. In the simplest model, one takes only the excluded
volume effects into account. A better model augments this description by
introducing some attractive inter-protein contacts. Here, we do it in a
different way than described in ref. \cite{airwater}, where 
the selection of possible interactions was based on the consideration
of the native contact map of one protein. Instead, we couple 
two hydrophobic residues, at site $i$ on one protein and at site $j'$
on another, whenever the distance between their $\alpha$-C atoms is
smaller than 12 {\AA}. We describe the coupling by the Lennard-Jones
potential in which the energy parameter is given by $\lambda \;\varepsilon$.
If $\lambda=1$ then the depth of the potential well is the same as for the
intra-protein contacts; if $\lambda=0$ then only the steric
repulsion is involved. 
The length parameter, $\sigma '$,
is set to $(R_{\alpha C, i} + R_{\alpha C, j'})/2$. Here, $R_{\alpha C,i}$
denotes the most likely radius of an effective sphere that
can be associated with the $i$th residue when it forms
the overlap-based contact that involves the residue. The values of
$R_{\alpha C,i}$ are residue-specific and are listed in ref. \cite{mcpoma}.\\

Another change is that we augment the single-protein simulational
geometry by introducing a repulsive square box in the $x-y$ plane. We release
$N_p$ identical proteins simultaneously, at random $x$, $y$ locations
near the bottom. 
The lateral size, $l$, of the box determines
the 2-dimensional number density, $n_2$ of the proteins that
arrive at the interface and stay there diffusing. We consider
$N_p$ of either 20 or 40
and $l$ that is initially set to 2000 nm and then adiabatically 
changed to 10 nm within 10 000 $\tau$. The system is then evolved
for an additional 100 000 $\tau$.
 The time scale of the simulations has been chosen based
on our previous studies of protein G and lysozyme \cite{airwater}.
$N_f$ of the $N_p$ proteins form a layer at the interface. The remaining
proteins cannot squeeze in and are found in the second and higher
order layers.\\

\begin{figure}[h]
\centering
\includegraphics[width=0.5\textwidth]{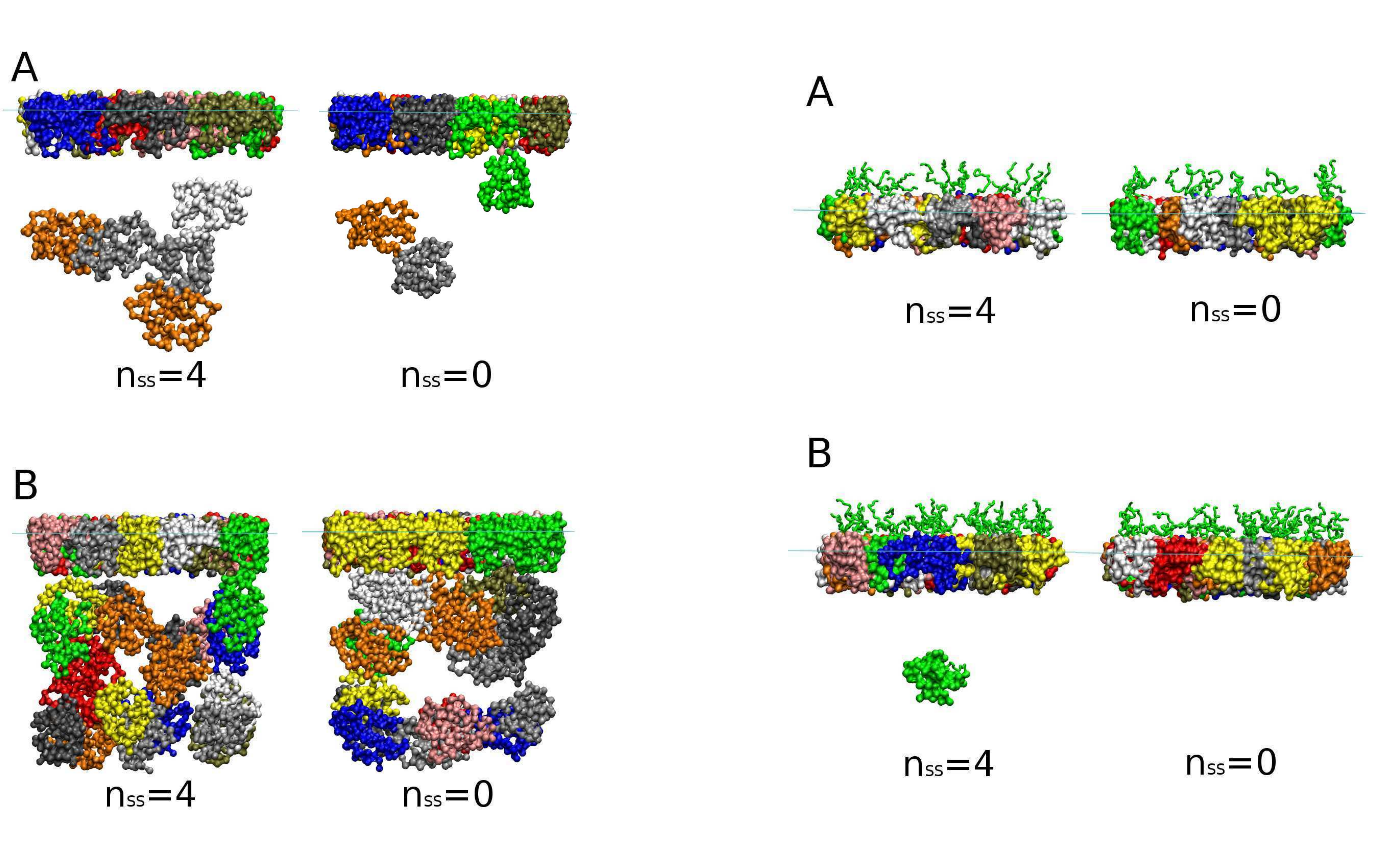}
\caption{Examples of conformations obtained at the air-water interface
for $N_p$ proteins with $\lambda=0$ -- the side views.
The left half of the figure is for  the LTP1 molecules and 
the right half -- for the LTP1b molecules (the ligands are in green). 
The left panels in each half
are for $n_{SS}$ of 4 and the right panels for $n_{SS}$ of 0.
The upper half of the figure is for $N_p$ of 20 and the lower half
for $N_p$ of 40. The corresponding numbers of the proteins
in the first layer are listed in Table \ref{n2group20}.
} \label{awsituation}
\end{figure}

We first consider the case of $\lambda=0$.
Table~\ref{n2group20} gives the values of $N_f$ in the case of $N_p=20$ 
and 40 at $T=0.3 \varepsilon/k_B$. The proteins are either of one kind
or mixed evenly -- the situation referred to as mixed.
We consider only the cases of $n_{SS}$=4 and 0.
The table indicates that $N_f$ is larger for LTP1b than for LTP1,
irrespective of the number of the  disulfide bonds. The smaller surface 
area that characterizes LTP1b at the interface allows for more molecules 
to be placed at the first layer, as illustrated in  Fig. \ref{awsituation}.
Moreover, the ligand adduct of LTP1b 
contributes to a better interface adsorption compared to  LTP1. 
As a result, $n_2$ in the case of LTP1b is 0.20 nm$^{-2}$ for $N_p=20$ 
and for the two considered values of $n_{SS}$, while in the case of LTP1 
it is 0.15 or 0.17 nm$^{-2}$ depending on whether $n_{SS}$ is 4 or 0. 
In the mixed case with $N_p=20$, all LTP1b molecules are adsorbed
in the first layer while some LTP1 are found in the second layer
($n_2$ is 0.18 ($n_{SS}=4$) or 0.19 nm$^{-2}$ ($n_{SS}=0$)).
All of this data indicates that LTP1b leads to denser protein layers.\\

\begin{figure}[h]
\centering
\includegraphics[width=0.5\textwidth]{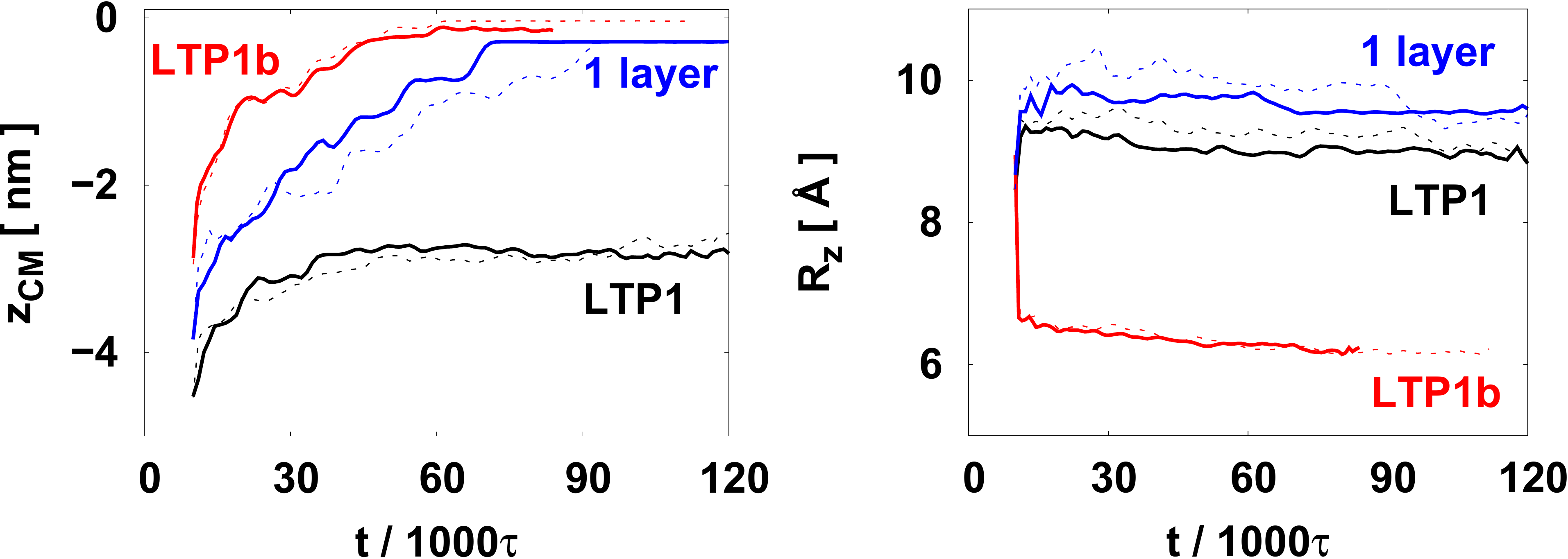}
\caption{The time evolution of the average $z_{CM}$ and $R_z$ for LTP1
(the black lines) and LTP1b (the red lines) in the case of $N_p$=40 with $\lambda=0$.
The solid lines correspond to $n_{SS}=4$ and the dashed lines to $n_{SS}=0$. 
The blue lines, labelled with "1 layer", correspond to the situation in which all
LTP1 proteins arrange into just one layer:  $N_p=15$ for $n_{SS}=4$ and 17 for $n_{SS}=0$. 
} \label{group40}
\end{figure}

Fig. \ref{group40} illustrates the dynamics of the proteins coming to 
the interface for $N_p$=40 ($\lambda=0$). The left panel shows the time dependence
of the average center of mass in the $z$ direction, $z_{CM}$. The right
panel shows the time dependence of the average $R_z$. It appears that the 
changes  in the geometry take place faster than the progression toward
the surface. The rate of progression is not sensitive to the value of
$n_{SS}$.\\

\begin{figure}[h]
\centering
\includegraphics[width=0.5\textwidth]{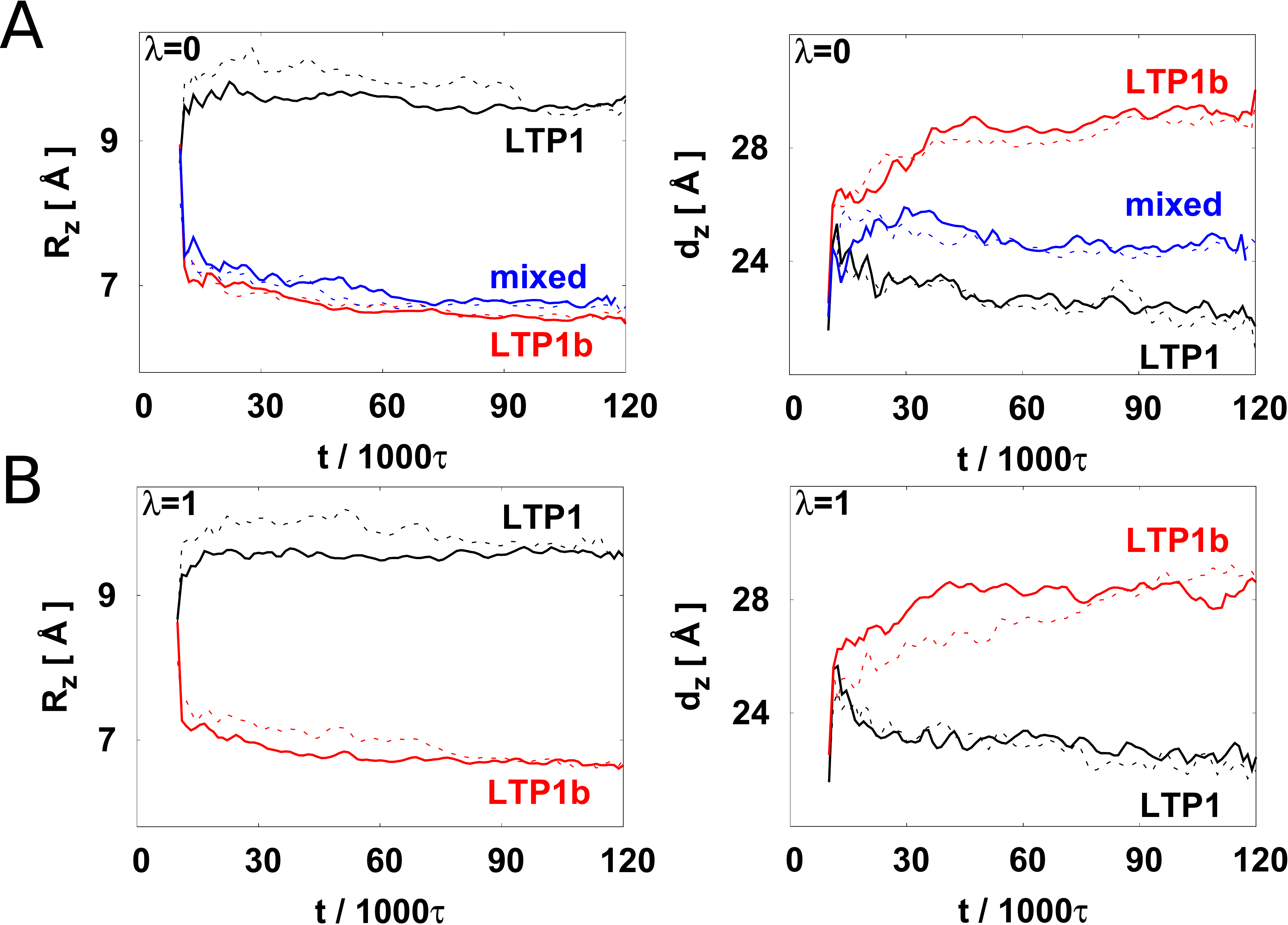}
\caption{A: The time evolution of the average $R_z$ and $d_z$
for LTP1 (the black lines), LTP1b (the red lines) and their mixture (the blue lines) 
in the case of $N_p$=20 with $\lambda=0$.
The solid lines correspond to $n_{SS}=4$ and the dashed lines to $n_{SS}=0$.
B: The same as A but for $\lambda=1$.
} \label{groupmix}
\end{figure}

The time-dependence of the shape parameters $R_z$ and $d_z$ for the 
LTP1, LTP1b and the mixed systems of $N_p=20$ is shown in Fig. \ref{groupmix}
for $n_{SS}$ of 0 and 4. For $\lambda=0$, $R_z$ of the mixed case
is much closer to that of LTP1b than LTP1, while $d_z$ is in between.
(The corresponding $R_z$ is about 30\% smaller for LTP1 but about 
3\% larger than for LTP1b;
$d_z$ is 9\% larger than for LTP1 but 13\% smaller than for LTP1b).\\

Fig. \ref{groupmix} (panels B) also addresses the situation with $\lambda=1$.
We observe that the effect of the attractive contacts on the shape
parameters is minor -- smaller than that associated with the
variations in $n_{SS}$. For LTP1, at the end of the time evolution,
$R_{z,\lambda=0,n_{SS}=4}$ is 0.4\% smaller than $R_{z,\lambda=1,n_{SS}=4}$
and 2.5\% smaller than $R_{z,\lambda=0,n_{SS}=0}$. 
Also, $d_{z,\lambda=0,n_{SS}=4}$ of LTP1 is 0.9\% larger than 
$d_{z,\lambda=1,n_{SS}=4}$ and 0.7\% larger than $d_{z,\lambda=0,n_{SS}=0}$. 
Similar results  are also obtained for LTP1b.\\

\subsection{The temperature dependence of the size of the cavity}

 
One problem arising when determining the volume of a protein cavity, which is
exposed to the solvent, is how to decide about its closure. We avoid making
such decisions by using the spaceball algorithm \cite{spaceball} which
relies on the statistical analysis of the volume calculations obtained for various
rotated grid orientations with respect to the protein.
We surround the protein by a rectangular box and generate a grid of 
points through intersections of lines that are parallel to the box edges.
The lines are taken to be separated by 0.2 {\AA}.
We then take spherical molecular probes (of radius 1.42 {\AA}) and
"walk" them along the three lattice directions until they encounter
an atom of the protein or the opposite wall.  The van der Waals radii of
the atoms are taken from ref. \cite{Tsai}. The sites that have not been
visited define the cavity space for this particular orientation. We then consider
25 rotations to change the orientation and average the results.
The average defines the most typical (as opposed to extremal) value
of the volume. As mentioned in the Introduction, this method yields
$V_c$ of 69.192 and 666.488 {\AA}$^3$ for LTP1 and LTP1b (with the
removed ligand) respectively, if one uses the PDB structure files. 
In the case of LTP1, there are
two disconnected cavities of 43.384 and 25.808 {\AA}$^3$. In the case of
LTP1b -- three of 262.784, 212.616, and 191.088 {\AA}$^3$.\\

It should be noted that different estimates have been obtained with the use of
the {\it VOIDOO} program \cite{voidoo}: no cavity in the case of LTP1 and
two cavities of 548 and 568 {\AA}$^3$ in the case of LTP1b \cite{bakan2009}.
In this program,  one first identifies the outside surface of the
protein and then sets a grid of lattice points that are surrounded
by this surface. One considers one orientation of the grid
and the grid spacing is set between 0.5
and 1.0 {\AA}. In order to take into account the excluded volume effects,
one uses a probe of radius 1.4 {\AA}. 
The grid points are all initialised to count as 0. This value is turned to 1 if
the grid point
distance to the closest protein atom is smaller than the sum of the 
probe radius and the van der Waals radius associated with the atom.
All grid points with the 0 value are away from the cavity wall
and thus count as contributing to the volume of the cavity.
In this procedure, the opening of the cavity is typically
ill-defined and one alleviates the problem by enlarging, or "fattening",
the van der Waals radii by a factor until the cavity gets closed.
The {\it VOIDOO} program  gives larger volumes than the procedure used
by us because the opening of the cavity counts too much
even with the fattening procedure.
We consider our procedure to be more accurate because our grid size is smaller
and because the enlargment of the radii to define the closure of the
cavity introduces errors also away from the closure. We have made an independent
check of the cavity geometry by indentifying the atoms on the inside of the cavity
and determining distances between them. By doing so we could determine that
LTP1 and LTP1b can accomodate about 3 and 7 water molecules respectively.\\

It is interesting to find out how does $V_c$ depend on $T$. We use the
NAMD all-atom molecular dynamics package \cite{NAMD} with the CHARMM22 \cite{charmm1}
force field to generate conformations corresponding
to a given $T$ and apply the spaceball algorithm to a sample of conformations.
The ASY ligand is removed from LTP1b. The system is equilibrated for 2 ns, 
in which time the temperature is increased 
in three steps from 110K, to 
210K, and to the final $T$ (of 250K, 300K, 325K, 350K, 375K and 400K). 
For $T\le 200$K, the system is  equilibrated in one step.
We pick 10 conformations for further analysis.\\

\begin{figure}[h]
\centering
\includegraphics[width=0.45\textwidth]{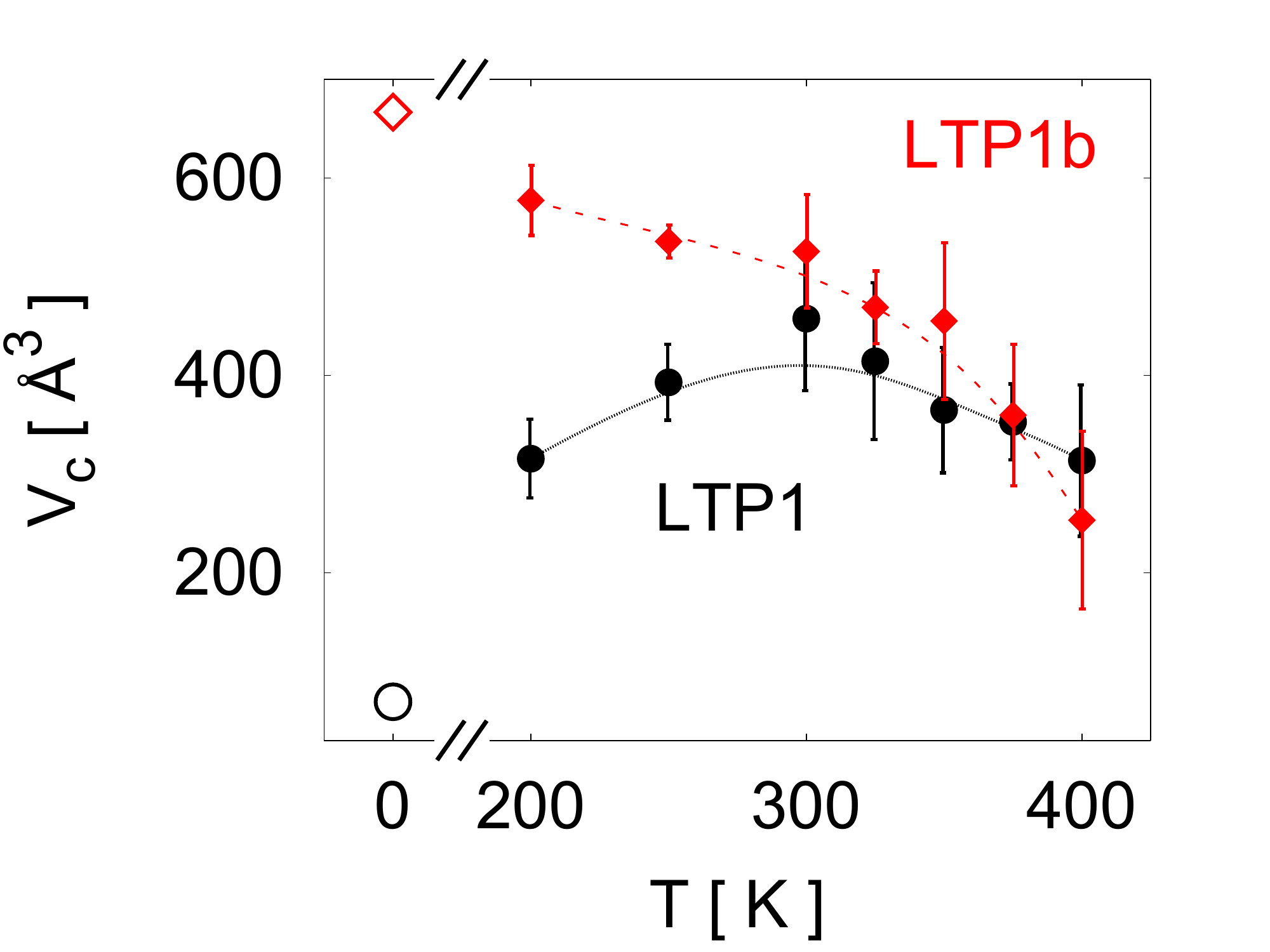}
\caption{The comparison of the cavity volume $V_c$ of LTP1 (black) and LTP1b (red, after
removing the ligand) at different temperatures. The data for $T=0$ K 
(the empty symbols) correponds to the native state of the protein.
All protein conformations are obtained from NAMD all-atom simulation in case of $n_{SS}=4$.
} \label{spaceb}
\end{figure}

Fig.~\ref{spaceb} shows the estimated $V_c$ for LTP1 and LTP1b as a function of $T$. 
For LTP1b, $V_c$ is seen to decrease monotonically as $T$ increases. 
This is because the stronger thermal fluctuations limit the free space inside of the
protein. For LTP1, $V_c$ increases from 69.192 {\AA}$^3$  
(see the empty black circle in Fig.~\ref{spaceb}) in the native state to 
457.564 {\AA}$^3$ at $T$=300K, then drops down monotonically as $T$ increases still further. 
The difference in behavior between LTP1b and LTP1 stems from
the fact that LTP1b is loosely packed, especially after removing the ligand,
whereas LTP1 is tightly packed. As $T$ increases, LTP1 gets partially unfolded
which makes the protein swallen and endowed with a bigger cavity. However,
on a further increase in $T$, the thermal fluctuations reduce the effective
volume of the cavity. For LTP1b, it is only the thermal fluctuations that
affect the volume of the cavity.
We observe that even though $V_c$ for LTP1b is larger than for LTP1 at room $T$,
the volumes  become more and more alike as $T$ grows. This is because of the 
smaller rigidity of LTP1b, as evidenced in Fig. \ref{folding}.\\

\section{Conclusions}

\begin{figure}[h]
\centering
\includegraphics[width=0.5\textwidth]{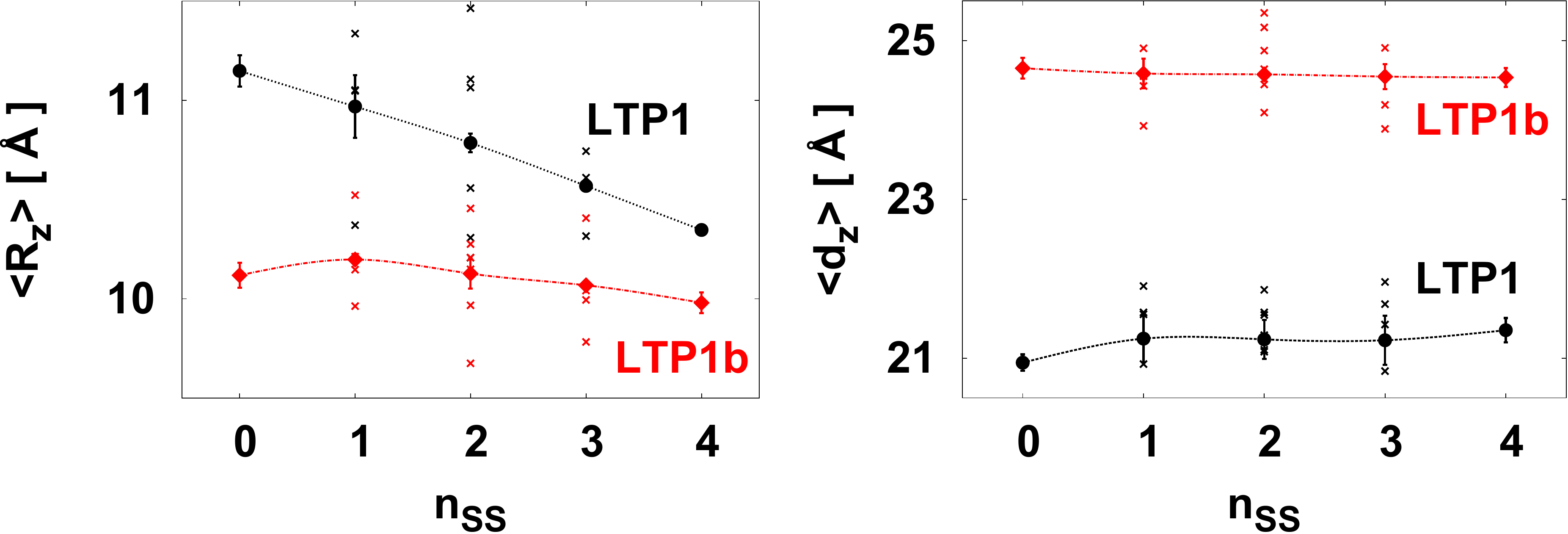}
\caption{$\langle R_z \rangle$ and $\langle d_z \rangle$ 
for LTP1 (black circles) and LTP1b (red diamonds) at $T=0.3 \varepsilon/k_B$
as a function of $n_{SS}$.
For $1\le n_{SS}\le 3$, the data of each permutation is displayed as crosses.  
The errors of the means have been obtained by considering 50, 80, 120, 80 and 50 
trajectories for $n_{SS}$ between 0 and 4 respectively and by partitioning the
data into groups.
} \label{shapex}
\end{figure}

\begin{figure}[h]
\centering
\includegraphics[width=0.5\textwidth]{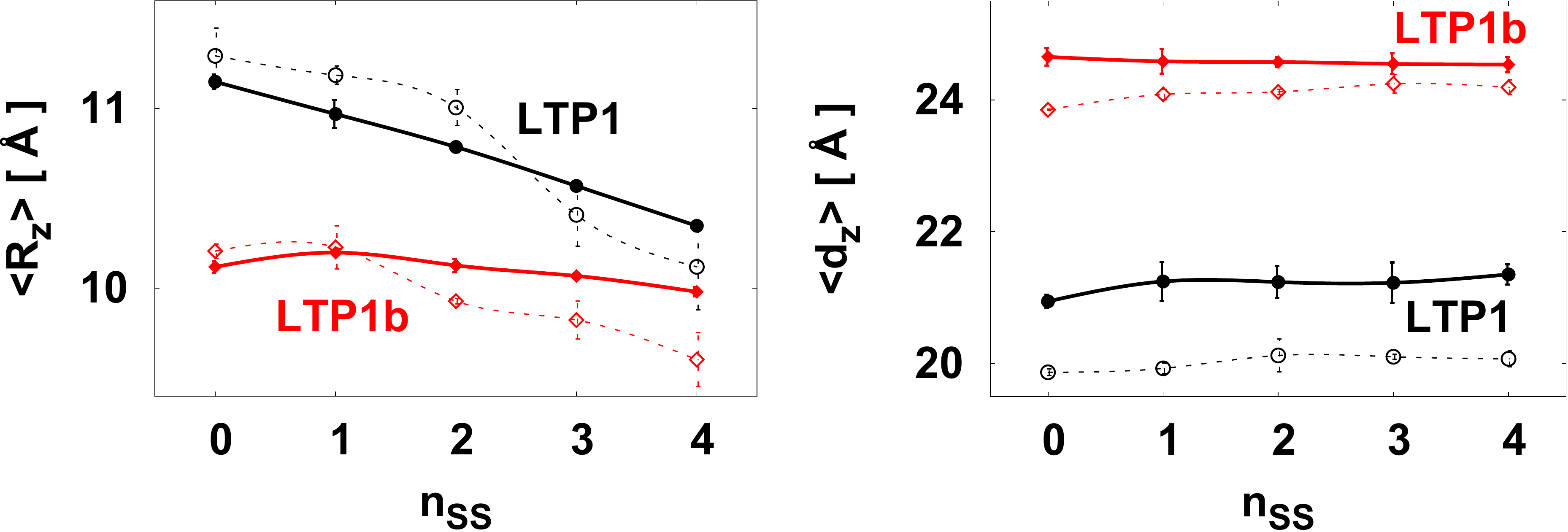}
\caption{The comparison of $<R_z>$ and $<d_z>$ between the glycated 
(empty data points) and non-glycated (full data points)
LTP1 and LTP1b for the 5 values of $n_{SS}$. The results are averaged
over the permutations in the placement of the disulfide bonds. 
The error bars have been estimated as in Fig. \ref{shapex}.
} \label{shapsuger}
\end{figure}

We have used the structure-based coarse-grained model to elucidate the
nature of the conformational transformations of LTP1 and LTP1b at the 
air-water interface in the context of beer foaming.
We have constructed an all-atom model that supports the
basics of the phenomenological description of the interface used in the
coarse-grained model.
Though our results are of a fairly qualitative nature, they
provide molecular-level insights into the process. Both of these
proteins are shown to deform and span the interface to stabilize it.
The degree of spreading depends on the number of the disulfide bonds: the 
smaller this number, the larger the surface area covered (see Fig.~\ref{shapex}).
We find that LTP1 spreads more than LTP1b because of the fatty
ligand.  The ligand makes the protein layer to be more packed and thicker.
The increased thickness should contribute to a slower flow rate of liquid 
drainage of beer foams observed by Bamforth et al. \cite{evan2009}.   
We also show that glycation increases the surface area at sufficiently
high levels of the disulfide-bond reduction.
We have argued that the \{3,50\} disulfide bond is more likely to be cleaved 
than \{48,87\} and that the structural properties of the proteins at the
interface depend on which bonds are actually cleaved.
We have provided new estimates of the volumes of the cavities
in the two proteins and showed that the volumes generally decrease
on heating. Thus heating should lead to a smaller propensity to
bind free lipids.\\

{\bf Acknowledgements}

We appreciate fruitful discussions with G. Rose and A. Sienkiewicz as well
the technical help from M. Chwastyk.
The project has been supported by
the National Science Centre (Poland),
Grant No. 2014/15/B/ST3/01905,
European Framework Programme VII NMP grant 604530-2
(CellulosomePlus).

\clearpage

\begin{table}[ht]
\caption{The list of possibilities of having the disulfide bonds in LTP1 and 
LTP1b under the conditions of various levels of reduction as characterized
by $0\le n_{SS}\le 4$. For $1\le n_{SS}\le 3$, the permutations are
denoted by $P_{n_{SS},{\lambda}}$ where $\lambda$ varies between
1 and 6 if $n_{SS}$=2 or between 1 and 4 if $n_{SS}$ is either 1 or 3.}
\centering
\begin{tabular}{|c |c |c |}
\hline
\multirow{1}{*}{$n_{SS}$} & number of permutation  & list of permutations $P_{n_{SS},{\lambda}}$\\
\hline
4 & 1 & \{3,50\} \{13,27\} \{28,73\} \{48,87\} \\
\hline
3 & 4 & \begin{tabular}{c}$P_{3,1}$: \{3,50\} \{13,27\} \{28,73\}\\$P_{3,2}$: \{3,50\} \{13,27\} \{48,87\}\\$P_{3,3}$: \{3,50\} \{28,73\} \{48,87\}\\$P_{3,4}$: \{13,27\} \{28,73\} \{48,87\} \end{tabular} \\
\hline
2 & 6 & \begin{tabular}{c}$P_{2,1}$: \{3,50\} \{13,27\}\\$P_{2,2}$: \{3,50\} \{28,73\} \\$P_{2,3}$: \{3,50\} \{48,87\}\\$P_{2,4}$: \{13,27\} \{28,73\}\\$P_{2,5}$: \{13,27\} \{48,87\}\\$P_{2,6}$: \{28,73\} \{48,87\}\end{tabular} \\
\hline
1 & 4 & \begin{tabular}{c}$P_{1,1}$: \{3,50\}\\$P_{1,2}$: \{13,27\}\\$P_{1,3}$: \{28,73\} \\$P_{1,4}$: \{48,87\}\end{tabular} \\
\hline
0 & 1 &  -- \\
\hline
\end{tabular}
\label{permuta}
\end{table}

\begin{table}[ht]
\caption{The average values of $R_z$, $d_z$ and $w$ for LTP1 and LTP1b,
together with their relative differences,
as a function of $n_{SS}$ at $T=0.3 \varepsilon/k_B$. 
For $n_{SS}=4$ or 0, the data is averaged over 20 trajectories. 
For $1\le n_{SS}\le 3$, we have generated 10 trajectories for each permutation.}
\centering
\begin{tabular}{|c |c |c |c |c |c |c | c|c|c|}
\hline
\multirow{2}{*}{$n_{SS}$} & \multicolumn{3}{c|}{LTP1}  & \multicolumn{3}{c|}{LTP1b} & \multirow{2}{*}{$r_{\langle R_z \rangle}$} & \multirow{2}{*}{$r_{\langle d_z \rangle}$} & \multirow{2}{*}{$r_{\langle w \rangle}$}\\
\cline{2-7}
& $\langle R_z \rangle$ \AA & $\langle d_z \rangle$ \AA & $\langle w \rangle$ & $\langle R_z \rangle$ \AA & $\langle d_z \rangle$ \AA & $\langle w \rangle$ & & &\\
\hline
4 & 10.15 & 21.35 &+0.02 & 9.98 & 24.54& -0.04 & -1.7\%& 14.9\%& -30.0\%\\
3 & 10.57 & 21.23 &-0.01 & 10.07 & 24.55& -0.05 & -4.7\%&15.6\%& -40.0\%\\
2 & 10.79 & 21.24 &-0.03 & 10.13 & 24.57& -0.02 & -6.1\%&15.7\%& 33.0\%\\
1 & 10.97 & 21.25 &-0.03 & 10.20 & 24.59& -0.04 & -7.0\%&15.7\%& -33.0\%\\
0 & 11.15 & 20.94 &-0.05 & 10.12 & 24.65& -0.02 & -9.2\%&17.7\%& 60.0\%\\
\hline
\end{tabular}
\label{compinf}
\end{table}

\begin{table}[ht]
\caption{The number of adsorbed molecules $N_{f}$ in the first layer
for the total number of $N_p$ proteins for $\lambda=0$ and for 
a given value of $n_{SS}$. 
The term mixed refers to a situation in which 50\% of the proteins
are LTP1 and the other 50\% are LTP1b. }
\centering
\begin{tabular}{|c |c |c |c |c |}
\hline
\multirow{2}{*}{protein} & \multicolumn{2}{c|}{$N_{f}$ ($N_p=20$)}  & \multicolumn{2}{c|}{$N_{f}$ ($N_p=40$)}\\
\cline{2-5}
& $n_{SS}=4$ & $n_{SS}=0$ & $n_{SS}=4$ & $n_{SS}=0$  \\
\hline
LTP1 & 15 & 17 & 21 & 22 \\
LTP1b & 20 & 20 & 39 & 40 \\
mixed & 18 & 19 & -- & -- \\
\hline
\end{tabular}
\label{n2group20}
\end{table}

\clearpage

\clearpage
\begin{figure}[h]
\centering
\includegraphics[width=0.5\textwidth]{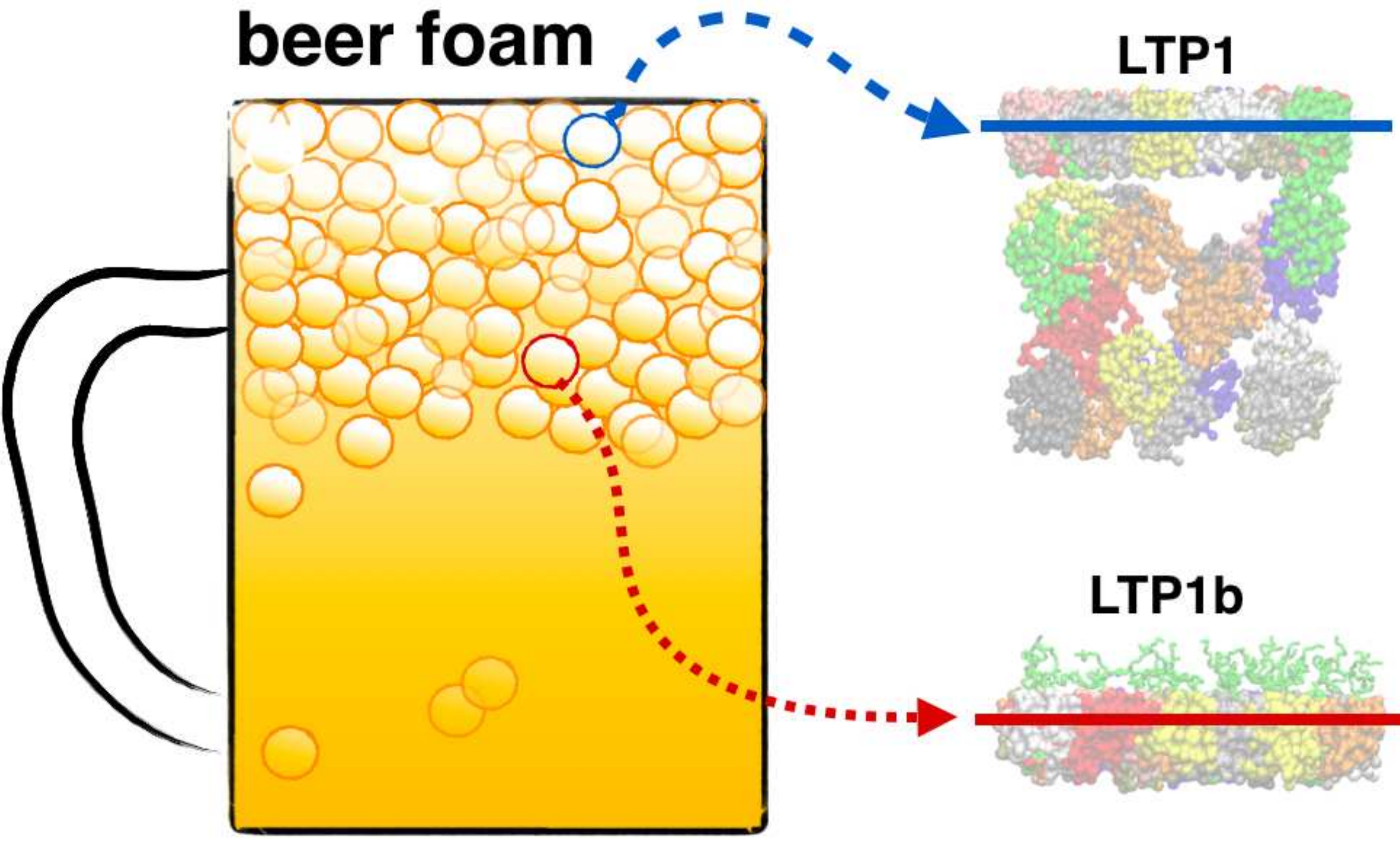}
\caption{The table of contents (TOC) graphic.
} \label{toc}
\end{figure}

\end{document}